\documentclass[fleqn,usenatbib]{mnras}

\usepackage[T1]{fontenc}
\usepackage{ae,aecompl}

\usepackage{graphicx}	
\usepackage{amsmath}	
\usepackage{amssymb}	
\usepackage[font=footnotesize,labelfont=bf]{caption}
\usepackage{xcolor}
\usepackage{gensymb}

\title[Neutron conversion-diffusion]{Neutron conversion-diffusion: a new model for structured short gamma-ray burst jets compatible with GRB 170817}

\author[E. Preau, K. Ioka and P. M\'esz\'aros]{
Edwan Preau,$^{1}$
Kunihito Ioka,$^{2}$
and Peter M\'esz\'aros$^{3,4,5,6}$
\\
$^{1}$Department of Physics, Ecole Normale Superieure, Paris, 75005, France\\
$^{2}$Center for Gravitational Physics, Yukawa Institute for Theoretical Physics, Kyoto University, Kyoto, 606-8502, Japan 
\\
$^{3}$Department of Physics, The Pennsylvania State University, University Park, Pennsylvania 16802, USA\\
$^{4}$Department of Astronomy \& Astrophysics, The Pennsylvania State University, University Park, Pennsylvania 16802, USA\\
$^{5}$Institute for Gravitation and the Cosmos, The Pennsylvania State University, University Park, Pennsylvania 16802, USA\\
$^{6}$Center for Multimessenger Astrophysics, The Pennsylvania State University, University Park, Pennsylvania 16802, USA\\  }

\date{Accepted XXX. Received YYY; in original form ZZZ}

\pubyear{2021}

\begin{document}
\label{firstpage}
\pagerange{\pageref{firstpage}--\pageref{lastpage}}
\maketitle

\begin{abstract}
We present a generic theoretical model for the structuring of a relativistic jet 
propagating through the ejecta of a binary neutron star merger event,
introducing the effects of the neutron conversion-diffusion, which provides a
baryon flux propagating transversely from the ejecta towards the jet axis. This 
results naturally in an increased baryon load structure of the outer jet
with the approximate isotropic energy distribution $E_{iso}(\theta) \propto \theta^{-4}$,
which is compatible with the first gravitational wave and short gamma-ray burst event GW170817/GRB 170817A observed at an off-axis angle of the jet.
\end{abstract}




\section{Introduction}

The observations of GRB 170817A/GW170817 both as a gamma-ray burst (GRB) source 
\citep{GRB170817A,GRB170817A_GBM,GRB170817A_INT} and as a gravitational wave (GW) source \citep{GW170817,GWEM170817} have provided, 
for the first time, substantial observational evidence for having detected a GRB jet  
off-axis.  The main argument for this is that the distance determination
based on the GW observations \citep{GW170817_H017} indicates that the inferred isotropic-equivalent luminosity
is several orders of magnitude lower than that for most other short 
gamma-ray bursts (sGRBs), although the 
GRB light-curve and spectrum are similar to those of a typical sGRB.
Typical sGRBs are known or inferred to be at much greater distances, and 
hence they are likely to have 
been detected close to on-axis, thus leading to much higher isotropic-equivalent luminosities.
A detailed off-axis jet interpretation has been presented by, e.g.
\citet{GRB170817A,Granot+17,LK18,IokaNakamura2017,IokaNakamura2019}. Alternative explanations for the low 
isotropic equivalent luminosity have also been considered by, e.g. 
\citet{Kasliwal+17,GottliebNakarPiranEtAl2017,Nakar+18}, mainly centering upon the hypothesis of the gamma-ray emission 
being due to the cocoon associated with the break-out of the 
sGRB jet from the dynamical ejecta following the merger of the neutron star binary
which is thought to give rise to the GW and sGRB event.
Note that the afterglow observations, i.e. superluminal motion in radio \citep{MooleyDellerGottliebEtAl2018,GhirlandaSalafiaParagiEtAl2018} and the closure relation between the spectral index and light curve slope \citep{afterglow2018,Mooley+18c,Lamb+19}, indicate that a relativistic jet is launched and successfully breaks out the merger ejecta \citep{NagakuraHotokezakaSekiguchiEtAl2014,Hamidani+20,2020arXiv200710690H}, although the origin of the gamma-ray emission is not settled yet.

The simplest considerations of an off-axis jet whose isotropic energy $E_{iso}$ and bulk Lorentz
factor $\Gamma$ has a top-hat structure, i.e. an abrupt cut-off at a certain angle 
$\theta_{0}$, lead to inconsistencies in the light of the observations of GRB 170817A, in particular the slowly-rising light curves of the afterglows \citep[e.g.][]{Mooley+18a}, 
confirming previous suspicions that realistic GRB jets cannot have a top-hat profile.
Indeed, structured jets with $E_{iso}(\theta)$ and $\Gamma(\theta)$ were theoretically considered
early on \citep{Meszaros+98angle,Zhang+02angle}, and have been shown to arise
naturally in hydrodynamical simulations \citep[e.g.][]{MacFadyen+99,MizutaIoka2013}. The physical reason 
why the hydrodynamical and more recently also MHD simulations \citep[e.g.][]{Tchekhovskoy+08,Kathirgamaraju+18,Fernandez+19,Gottlieb2020} 
show such jet structuring
is the same for both long GRB jets emerging through a stellar envelope and
for short GRB jets emerging through a dynamical debris outflow: the light, highly 
relativistic jet interacts, at its outer edges, with a much slower, baryon-loaded 
outer material which ``pulls down" at the jet outer edges, resulting in a jet with
a faster inner jet core and an increasingly slower outer sheath. This is a purely
macroscopical mechanism, which however does not take generally into account baryon
composition (electron fraction) and microphysical effects such as diffusion \citep{LevinsonEichler2003}.

The observations of GRB 170817A afterglows imply that,
in terms of a structured jet which is observed off-axis,
the angular energy distribution may be
Gaussian \citep[e.g.][]{afterglow2018,Margutti+18,Lamb+19},
power-law \citep[e.g.][]{D'Avanzo+18,GhirlandaSalafiaParagiEtAl2018},
hollow-cone \citep{Takahashi+19,Nathanail+20} or spindle \citep{takahashi2020diverse}.
These afterglow observations only constrain
the inner jet structure near the core
\citep{IokaNakamura2019,Takahashi+19},
while the outer jet structure,
which is more relevant to the observed gamma-ray emission,
is not constrained well.
Specific hydrodynamic or MHD simulations of an sGRB jet moving
through the slow merger ejecta have so far not been detailed enough to
reliably obtain these angular structures.

In the present paper we present a theoretical model,
based on a microphysical effect, namely,
neutrons diffusing from the outer ejecta into the jet,
which successively convert into protons 
and back into neutrons through inelastic collisions,
diffusing towards the jet axis.
This results in an increased baryon load of the jet,
greater in the outer regions than in the inner regions,
which naturally gives rise to
an isotropic energy structure $E_{iso}(\theta)\propto \theta^{-4}$
for the outer jet.
We discuss the cases where the neutron conversion-diffusion can be essential for an interpretation of the gamma-ray observations of sGRB 170817A.

In \S 2 we present the basic model and its parameters. In \S 3 we estimate the total number of neutrons picked-up by the jet,
while in \S 4 we discuss in more details the diffusion of neutrons in the jet, and in \S 5 the resulting jet structure and the implications for sGRB 170817A.

We use the usual notation for a quantity $Q$ in CGS units $Q_{,x} = Q/10^x$.

\section{The system}

\begin{figure}
\begin{center}
\includegraphics[scale = 0.3]{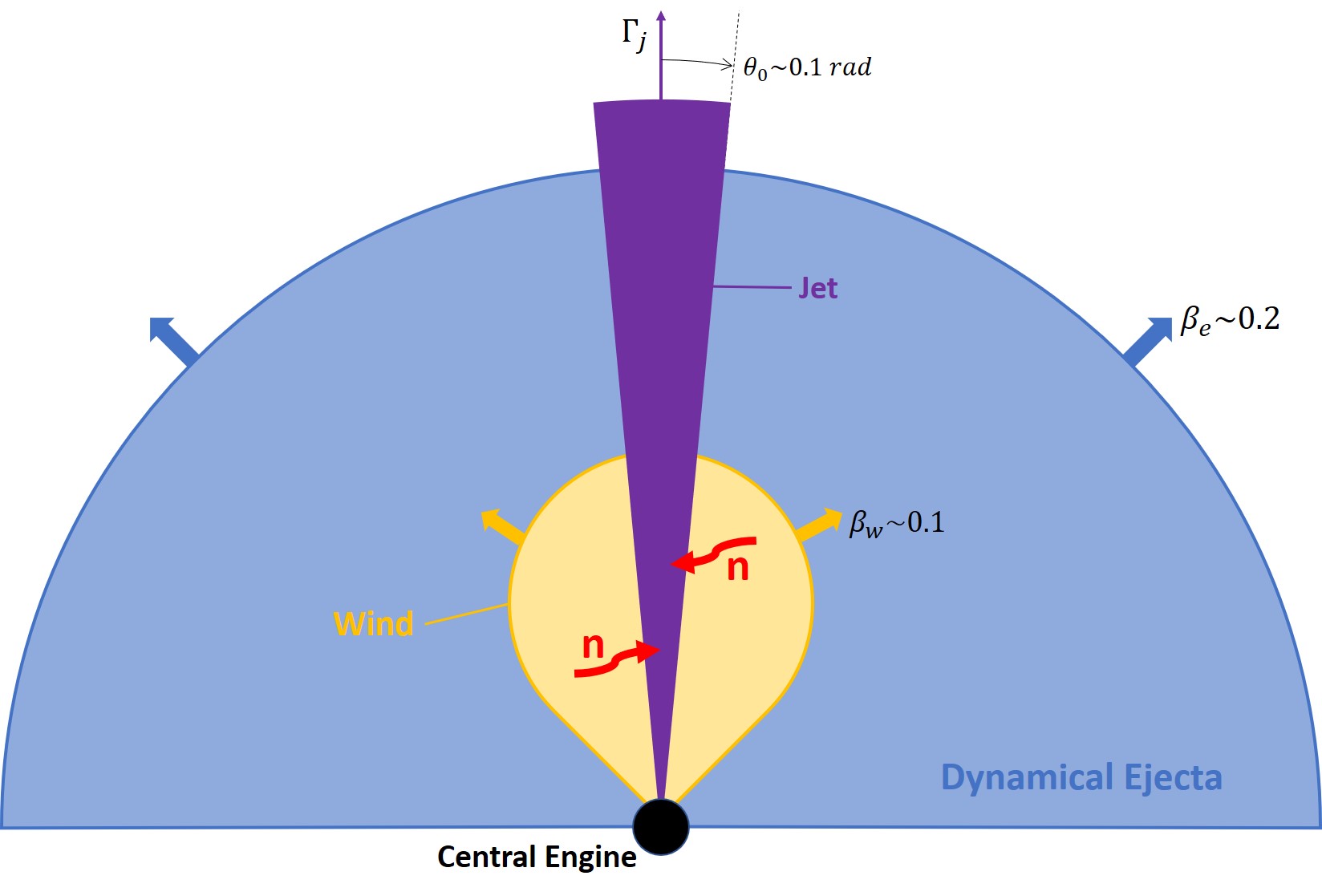}
\end{center}
\caption{Schematic representation of the system under consideration, consisting of the central engine (black sphere), the dynamical ejecta (blue sphere) and the wind/outflow (light orange). As the jet (in purple) propagates through the wind, it acquires baryons loaded through its outer edges via neutron leakage. The jet has a Lorentz factor $\Gamma_j$ and opening angle $\theta_0$. Note that the cocoon - the component produced by the jet interacting with the wind and the dynamical ejecta - is not shown explicitly. We argue later (Sec.~\ref{sec:pickup}) that it is a reasonable approximation to forget about the cocoon as far as the neutron leakage is concerned. }
\label{fig:sys}
\end{figure}

The configuration we consider is illustrated in Fig. \ref{fig:sys}. At time $t_0$ two neutron stars merge into a central engine that can be either a black hole or a hyper massive neutron star (HMNS). In the process, some portion of the total mass is ejected in the form of an ejecta. First a dynamical ejecta is emitted by shock and tidal effects during the merger, and a wind or outflow (hereafter wind) follows, originating from the accretion disk after a time of the order of the viscous timescale. At some time $t_j$ after the merger, a bayon-poor jet is launched from near the central engine \citep{Paczynski1986, Goodman1986, EichlerLivioPiranEtAl1989, IokaNakamura2017}. 

The jet needs to break through the surrounding ejecta before expanding freely in the low-density ISM to emit gamma-ray bursts \citep{NagakuraHotokezakaSekiguchiEtAl2014}. Whereas protons are forced to follow magnetic lines and therefore cannot move much in the ejecta, neutrons are free to drift in from the outside into the baryon poor jet \citep{MeszarosRees2000, LevinsonEichler2003}. As the jet propagates through the ejecta, it gets baryon-loaded starting at its edges by neutron pick-up through inelastic scattering with protons of the jet. The incoming neutron is converted by the inelastic collision into a proton that is then picked up along the magnetic field of the jet's flow. Note that this process has not been considered yet in hydrodynamical or MHD simulations. \\
\indent The sub-relativistic ejecta is made up of three components: 
\begin{itemize}

\item \underline{The dynamical ejecta:} we consider a mass $M_{e} \sim 0.01M_{\odot}$ expanding quasi-spherically \citep{Hotokezaka+13b} during the merger, caused by tidal and shock effects, which constitutes the dynamical ejecta. 

\item \underline{The neutrino-driven wind:} we assume that this is emitted conically within an injection angle $\psi \sim 1$ at a radius $r_{0} \sim 10^7$ cm (defined as the radius where the kinetic energy flux equals the thermal energy flux in the wind), a time

\begin{align}
t_{\nu} &\sim 0.07 \; s \left(\frac{M_{ns}}{2.5M_{\odot}}\right)r_{0,7}\left(\frac{L_{\nu ,e}}{3\times 10^{52} \; {\rm erg}\,{\rm s}^{-1}}\right)^{-1}
\nonumber\\
&\times
\left(\frac{\xi}{1.5}\right)^{-1}\left(\frac{E_{\nu,disc}}{15 \; {\rm MeV}}\right)^{-2},
\end{align}
after the merger \citep{PeregoRosswogCabezonEtAl2014}, where $M_{ns}$ is the mass of the HMNS, $L_{\nu,e}$ is the total power emitted in electron neutrinos, $\xi = (1 - \cos(\psi))^{-1}$ and $E_{\nu, disk}$ is the typical energy of neutrinos emitted from the disc. 

\item \underline{The viscous outflow:} The part of the accretion disk that expands due to viscous heating is referred to as the viscous outflow. It also expands out of the polar axis on viscous timescales $t > t_{disk}$, where the accretion time is given by \citep{PeregoRosswogCabezonEtAl2014}:

\begin{equation}
t_{disk} \sim 0.3 \; s \left(\frac{\alpha}{0.05}\right)^{-1} \left(\frac{H/R_{disc}}{1/3}\right)^{-2}\left(\frac{M_{ns}}{2.5M_{\odot}}\right)^{-1/2}R_{disk,7}^{3/2},
\end{equation}
with $\alpha$ the parameter for the viscosity of the disk ($\alpha$-disk model), $H$ the height of the disk along the polar axis and $R_{disk}$ its radius. The viscous outflow is expected to be the dominant ejecta as far as the total ejected mass is concerned \citep{SiegelMetzger112017, ShibataKiuchiSekiguchi2017, FujibayashiKiuchiNishimuraEtAl2017,Fujibayashi+20}.

\item \underline{The combined wind near the polar axis:} We model the ejecta near the polar axis by a single wind emitted a time $t_{w}$ after the merger with an opening angle $\psi \sim 1 \; \mathrm{rad}$ from a radius $r_{0} \sim 10^7 \; \mathrm{cm}$. This wind models the joint contributions of the neutrino-driven wind and the viscous outflow. Unlike the dynamical ejecta, this matter is continuously injected at the base of the wind. After the dynamical ejecta is \textit{r}-processed (about $1\, \mathrm{s}$ after the merger \citep{MetzgerMartinez-PinedoDarbhaEtAl2010}), the only place where neutrons exist significantly is at the base of the wind. We therefore neglect the dynamical ejecta as far as neutron leakage is concerned and turn our attention to the wind's dynamics.  
\end{itemize}

\subsection{The wind dynamics}

From $r_0$, we assume the velocity $v_{w} = \beta_{w} c \sim 0.1 \, c$ is constant. The neutron injection process is also assumed to be stationary, so that both $r_{0}$ and the power injected at $r_{0}$, $L_{w}$ (total two-sided power), are constant. 

The assumption that the velocity is constant is an approximation that amounts to neglecting the transition region where thermal energy is transferred to kinetic energy. Because of radioactive heating, this region is extended until \textit{r}-process ends. However, this occurs mainly in the range $r_0 < r < 10 \, r_{0}$ (during the formation of $\alpha$-particles which is the dominant heating process), and we will see that the baryon loading depends dominantly on the outer radii part of the integral. Therefore,  considering only the final value of the velocity (when $\beta_{w}(r)$ appears in the integral - see later) should yield the correct order of magnitude of the baryon loading.

\subsubsection*{Baryon density}

The conservation of baryon number as the wind expands radially gives:

\begin{equation}
n_b(r) = n_{b,0} \left(\frac{r}{r_0}\right)^{-2},
\end{equation}
where the baryon density at radius $r_0$, $n_{b,0}$, is obtained by the conservation of the baryon flux:

\begin{equation}
m_{p} n_{b,0} \beta_{w}c = \frac{\dot{M}}{4\pi r_{0}^2} \xi,
\end{equation}
$\dot{M}$ is the total (two-sided) mass flux ejected into the wind, whose fiducial value is taken so that a mass $\sim 0.005 \, M_{\odot}$ is ejected within $\sim 100 \; \mathrm{ms}$ (to fit with the simulation of the neutrino-driven wind in \citet{PeregoRosswogCabezonEtAl2014}): $\dot{M} \sim 10^{32} \; \mathrm{g\,s}^{-1}$. As a result, the density in the wind is expressed as:

\begin{equation}
n_{b}(r) \sim  2\times10^{31} \; \dot{M}_{32} r_{7}^{-2} \left(\frac{\xi}{1.5}\right) \left(\frac{\beta_{w}}{0.1}\right)^{-1} \mathrm{cm}^{-3}\, .
\end{equation}

\subsubsection*{Temperature}

As for the temperature, it is given by the conservation equation of the entropy. Note that for the radii we are interested in - where the wind is not \textit{r}-processed yet - the expansion is actually not adiabatic because entropy is injected by the \textit{r}-process (and formation of smaller nuclei before that, $\alpha$-process in particular). We define $r_{r-pro}$ the radius beyond which the wind is \textit{r}-processed.\\
\indent For $r_{0} < r < r_{r-pro}$, heat is modeled by a constant power per unit mass $\dot{Q}_{0} \sim 10^{19} \mathrm{erg\,g^{-1}\,s^{-1}}$ \citep{MetzgerMartinez-PinedoDarbhaEtAl2010}. The equation for the entropy evolution of a shell of wind with radial thickness $\delta r$ in the observer frame expanding radially then reads:

\begin{equation}
\frac{\mathrm{d}S}{\mathrm{d}r} = \frac{1}{\beta_{w}c} \frac{\dot{Q}_{0}}{T} n_{b}(r)m_{b} \; 2\pi \xi^{-1} r^2 \delta r,
\end{equation}
where $S = \frac{4a}{3}T^3 \; 2\pi \xi^{-1} r^2 \delta r$ is the entropy of the shell (the entropy of the wind is dominated by radiation), $m_{b}$ is the mass of a baryon and $a = 4\sigma/c$ with $\sigma$ the Stefan-Boltzmann constant.\\
\indent Then the temperature in the region where the wind is not \textit{r}-processed yet is:

\begin{equation}
T_w(r)^4 = T_{0}^{4}\left(\frac{r}{r_{0}}\right)^{-8/3} + T_{Q}^{4} \left[\left(\frac{r}{r_{0}}\right)^{-1} - \left(\frac{r}{r_{0}}\right)^{-8/3} \right],
\end{equation}
where $T_{Q}^4 = \frac{3}{5a}\dot{Q}_{0}\frac{r_{0}}{\beta_{w}c} n_{b,0}m_{b}$, so that:

\begin{equation}
\label{TQ} T_{Q} \sim 3\times10^9 \, \mathrm{K} \; \dot{M}_{32}^{1/4} r_{0,7}^{-1/4} \left(\frac{\xi}{1.5}\right)^{1/4} \left(\frac{\beta_{w}}{0.1}\right)^{-1/2} \dot{Q}_{0,19}^{1/4}. 
\end{equation}
Remember that $r_{0}$ is defined as the radius where kinetic and thermal energy densities are equal in the wind, and the baryon flux is conserved:

\begin{equation}
aT_{0}^4 \beta_{w}c \sim \frac{1}{2}m_{p} n_{b,0} (\beta_{w}c)^3 \sim \frac{1}{2} \frac{\dot{M}}{4\pi r_{0}^2} \xi (\beta_{w}c)^2,
\end{equation}
which gives $T_{0}$:

\begin{equation}
T_{0} \sim 10^{10} \mathrm{K}\; \dot{M}_{32}^{1/4}r_{0,7}^{-1/2} \left(\frac{\xi}{1.5}\right)^{1/4} \left(\frac{\beta_{w}}{0.1}\right)^{1/4}.
\end{equation}
For $r/r_{0} > (T_{0}/T_{Q})^{12/5} \sim 20 \; \dot{Q}_{0,19}^{-3/5}r_{0,7}^{-3/5}(\beta_{w}/0.1)^{9/5}$, the following approximation is therefore relevant:

\begin{equation}
\label{TwQ}T_{w}(r) \sim T_{Q} \left(\frac{r}{r_{0}}\right)^{-1/4}. 
\end{equation}
Once again, as we will see that the outer radii dominate the neutron diffusion in the integral, we will use the expression \eqref{TwQ} to describe the radial structure of the wind temperature.

\subsubsection*{\textit{R}-process}
The radius where the \textit{r}-process ends is crucial to the baryon loading of the jet by neutron leakage from the wind, because the neutron density drops dramatically above this radius. The \textit{r}-process ends after a time $\Delta t_{r-pro} \sim 0.1-1 \; \mathrm{s}$ \citep{MetzgerMartinez-PinedoDarbhaEtAl2010, MartinPeregoArconesEtAl2015}. Neglecting the $\alpha$-process ignition radius, we find:

\begin{equation}
\label{yrpro}r_{r-pro,7} \sim \frac{c\beta_{w} \Delta t_{r-pro}}{10^7\,{\rm cm}} \sim 60 \;  \left(\frac{\beta_{w}}{0.1}\right) \left(\frac{\Delta t_{r-pro}}{0.2\,{\rm s}}\right) \, .
\end{equation}
From now on we define the dimensionless radius

\begin{equation}
y \equiv \frac{r}{r_0} \, . 
\end{equation}

\section{Neutron pick-up}\label{sec:pickup}

In this section, we compute the total number of baryons picked-up by a shell of the jet with radial thickness $\delta r$ (in the central engine frame) through neutron leakage from the wind. This neutron injection is assumed, for simplicity, to be stationary. The shell is emitted from the origin at $t = t_{j} + \Delta t$ where $t_j$ is the time after the merger when the jet launch starts. We take $\Delta t$ as the delay time of the shell behind the tip of the jet.
 
For simplicity, we will also assume that $\Delta t > \Delta t_r$, with

\begin{equation}
\label{defDr}\Delta t_r = \frac{(y_{r-pro} - 1)(1 - \beta_{w})r_0/c}{\beta_{w}} \sim 0.3 \; \mathrm{s} \; \left(\frac{y_{r-pro}}{100}\right)r_{0,7}\left(\frac{\beta_{w}}{0.1}\right)^{-1}.
\end{equation}
This condition ensures that $1 + \frac{\beta_{w}}{1 - \beta_{w}} \frac{\Delta t}{r_0/c} > y_{r-pro}$, so that every part of the wind relevant to neutron leakage has been emitted after $t_j$. This will turn out to be very convenient when evaluating the neutron diffusion time. 

The jet advance is accompanied by the formation of a semi-relativistic cocoon which surrounds it. This cocoon also expands beyond $y_{r-pro}$ for such $\Delta t$ (as it is advected with the wind). This may bring an additional contribution to the neutron leakage for radii bigger than $y_{r-pro}$ if the cocoon (composed of wind and jet gas) is not \textit{r}-processed yet (which may be possible even beyond $y_{r-pro}$ because it has been heated-up by energy injection from the jet), but this contribution should be negligible due to the cocoon expansion into the dynamical ejecta. To make things simple, we also suppose $\Delta t$ is such that the cocoon is already \textit{r}-processed. \\
\indent Therefore, we consider that neutron leakage happens exclusively between $r_0$ and $r_{r-pro}$, directly from the wind.

\subsection{Neutron diffusion in the wind}

The number of picked-up neutrons, as the shell moves a distance $\mathrm{d}r$ is :

\begin{equation}
\label{defddNpu}\mathrm{d}\delta N_{pu}(r, \Delta t) = J_{D}(r, \Delta t) \times 2\pi r\sin{\theta_{0}}\delta r \frac{\mathrm{d}r}{c}.
\end{equation}
That is the flux of neutrons $J_{D}$ multiplied by the infinitesimal external surface of the interface between the shell and the wind, and the time to go a distance $\mathrm{d}r$. Here $\theta_{0}$ is the opening angle of the jet. \\
\indent The neutron flux is given by the diffusive flux maintained by the elastic scattering of neutrons on protons in the wind, that is, by a diffusion process where the jet is equivalent to an absorbing wall \citep{LevinsonEichler2003}: 

\begin{equation}
J_{D} = K\frac{\partial n_{n,w}}{\partial x}(x = x_{jet}) \;\;\;\;\; , \;\;\;\;\; K = \lambda_{np} v_{th},
\end{equation}
where $K$ is the diffusion coefficient, $\lambda_{np} = 1/(n_{p,w}\sigma_{el})$ is the mean free path for a neutron colliding elastically with protons and $v_{th} \sim (3k_BT_w/m_n)^{1/2}$ is the thermal speed of neutrons in the wind. Here, $n_{n,w}$, $n_{p,w}$ and $\sigma_{el} \sim 30 \, \mathrm{mbarn}\times (c/v_{th}) \sim 1 \; \mathrm{barn}$ are the neutron density in the wind, the proton density in the wind and the cross-section for elastic scattering, respectively, and $x$ is the cylindrical radius, that is the distance from the jet axis. 

Solving the diffusion problem gives a relation between the flux of neutrons and the density gradient length scale $l$:

\begin{equation}
J_{D} \sim \frac{1}{2}\lambda_{np}v_{th}\frac{n_{n,w}(x = \infty )}{l},
\end{equation}
where $l$ is related to the time $\Delta t_{diff}(r)$ since neutrons at radius $r$ started leaking into the jet - that is the time since the corresponding radius has been reached by the jet:

\begin{equation}
l = (\lambda_{np}v_{th}\Delta t_{diff})^{1/2}.
\end{equation}
Note that the wind has a typical electron fraction $Y_{e} \sim 0.2 - 0.4$ (for the neutrino-driven wind see \citet{MartinPeregoArconesEtAl2015}; for the viscous outflow see \citet{SiegelMetzger112017}). The corresponding expression for the neutron flux is then given by:

\begin{align}
J_{D} &= \frac{1}{2}\Delta t_{diff}^{-1/2} \sqrt{\lambda_{np}v_{th}}(1 - Y_{e})n_{b}\, , \\
\label{JD} &\sim 10^{32} \; \left(\frac{r}{r_{0}}\right)^{-9/8} \frac{1 - Y_{e}}{2\sqrt{Y_{e}}} \Delta t_{diff}^{-1/2} \dot{M}_{32}^{5/8} r_{0,7}^{-9/8} 
\nonumber\\
&\times  \left(\frac{\xi}{1.5}\right)^{5/8}  \left(\frac{\beta_{w}}{0.1}\right)^{-3/4}\dot{Q}_{0,19}^{1/8} \; \mathrm{cm^{-2}\,s^{-1}}.
\end{align}
Note that we assumed that every incoming neutron is picked-up, which is consistent with the estimate for the mean free path of an incoming neutron before it collides inelastically with a proton from the jet.

\subsection{Baryon loading}

For $\Delta t > \Delta t_r$ and $r_0 < r < r_{r-pro}$, the diffusion time $\Delta t_{diff}$ depends on the radius as:

\begin{equation}
\label{Dtdiff}\Delta t_{diff} = (y - 1) \frac{r_0/c}{\beta_{w}}.
\end{equation}
Then the infinitesimal number of baryons picked-up at each step $\mathrm{d}r$ is, by substituting Eq. \eqref{JD} into Eq. \eqref{defddNpu}:

\begin{align}
\label{ddNpu}\mathrm{d}\delta N_{pu}(y) & \sim 3\times10^{43} \; \delta y \mathrm{d}y\; y^{-1/8}(y - 1)^{-1/2} \theta_{0,-1}\\
\nonumber &\times r_{0,7}^{13/8} \frac{1 - Y_{e}}{2\sqrt{Y_{e}}} \dot{M}_{32}^{5/8} \left(\frac{\xi}{1.5}\right)^{5/8} \left(\frac{\beta_{w}}{0.1}\right)^{-1/4}  \dot{Q}_{0,19}^{1/8}.
\end{align}
To compute the full baryon loading from neutron leakage of the shell, we now integrate from the radius of injection ($y=1$) to the radius where neutron leakage becomes negligible, that is the \textit{r}-process radius: 

\begin{equation}
\delta N_{pu} = \int_{1}^{y_{r-pro}}\mathrm{d}\delta N_{pu}(y).
\end{equation}
Substituting Eq. \eqref{ddNpu}:

\begin{align}
\delta N_{pu} &\sim 6 \times 10^{43} \; \delta y\, y_{r-pro}^{-1/8} \sqrt{y_{r-pro} - 1} \\
 \nonumber &\times \theta_{0,-1}r_{0,7}^{13/8} \frac{1 - Y_{e}}{2\sqrt{Y_{e}}} \dot{M}_{32}^{5/8} \left(\frac{\xi}{1.5}\right)^{5/8}  \left(\frac{\beta_{w}}{0.1}\right)^{-1/4}\dot{Q}_{0,19}^{1/8} ,
\end{align}
where we approximated $y^{-1/8}$ by its value for the highest radii $y^{-1/8} \sim y_{r-pro}^{-1/8}$ to evaluate the integral. With Eq. \eqref{yrpro} we find:

\begin{align}
\label{dNpuref} \delta N_{pu} &\sim 4\times10^{44} \; \delta y \,\, \theta_{0,-1}r_{0,7}\\
\nonumber &\times \frac{1 - Y_{e}}{2\sqrt{Y_{e}}} \dot{M}_{32}^{5/8} \left(\frac{\xi}{1.5}\right)^{5/8}  \left(\frac{\beta_{w}}{0.1}\right)^{1/8} \dot{Q}_{0,19}^{1/8} \left(\frac{\Delta t_{r-pro}}{0.2 s}\right)^{3/8}  .
\end{align}
We will take this value - which is the stationary value for every shell of the jet after $\Delta t > \Delta t_{r}$ - as the fiducial value for the baryon loading of the jet by neutron pick-up from the wind.

\section{Diffusion in the jet}

The baryon diffusion towards the axis of the jet is crucial for the final jet structure that arises as a consequence of the neutron leakage. In this section, we first show that elastic scattering has a short mean free path affecting only the rim of the jet. Then we take into account the conversion of protons into neutrons via inelastic scattering, the so-called conversion mechanism \citep{DerishevAharonianKocharovskyEtAl2003}, which extends the effective diffusion length affecting the jet structure.  Since this is the first time that the conversion mechanism is explicitly and quantitatively combined with the diffusion, we coin the term conversion-diffusion model.

\subsection{Elastic scattering}

\label{eldif}

We temporarily assume that the neutrons which are picked-up in the jet after having been converted into a proton through inelastic scattering with a proton from the jet,

\begin{equation}
\label{npinel1} n + p \to p + p + \pi^{-} \to p + p + \mu^{-} + \bar{\nu}_{\mu},
\end{equation}
cannot further scatter towards the axis of the jet. These protons accumulate within the first mean free path and their added mass results in  a deceleration (a decrease of the radial bulk Lorentz factor $\Gamma$) of that portion of the jet, so that here the elastic scattering dominates the inelastic one. The neutrons then diffuse elastically inside the decelerated part. Once they reach the inner portion of the jet that is not baryon-loaded yet, they get picked-up by inelastic scattering with a proton from the jet. To estimate the thickness of the pick-up ring (the part of the jet where the incoming neutrons are picked up), we therefore need to compare the time required for neutrons to elastically scatter across the ring with the dynamical time. 

\subsubsection*{Inelastic mean free path} 

We first discuss the dependence of the proton-neutron collision cross-section on the kinetic energy in the neutron inertial frame. For kinetic energies superior to $\sim 1 \mathrm{GeV}$, the elastic cross-section is $\sigma_{el} \sim 10 \; \mathrm{mbarn}$ and the inelastic cross-section is $\sigma_{inel} \sim 30 \; \mathrm{mbarn}$ \citep{TanabashiHagiwaraHikasaEtAl2018}.\\
\indent The typical Lorentz factor (of the relativistic proton) under which inelastic collision becomes negligible is the pion production threshold Lorentz factor $\gamma_{\pi^-}$, given by:

\begin{equation}
s^{2} = -(m_{p} + m_{n} + m_{\pi^{-}})^2 = -m_{p}^2 - m_{n}^2 - 2\gamma_{\pi^{-}}m_{p}m_{n},
\end{equation} 
and hence
\begin{equation}
\label{gp}\gamma_{\pi^-} -1 \sim 0.31 ,
\end{equation}
with $s$ the usual Mandelstam variable. As for the Lorentz factor $\gamma_{el}$ under which the initial collision between an incoming neutron and protons from the jet is more likely to be elastic rather than inelastic, it is given by experimental data \citep{TanabashiHagiwaraHikasaEtAl2018}:

\begin{equation}
\sqrt{-s^2} \sim 2.5 \, \mathrm{GeV} \sim 1.35 \times (m_{p} + m_{n})c^2,
\end{equation} 
with 

\begin{equation}
s^2 = (p_n + p_p)^2 = -m_{n}^2 - m_{p}^2 - 2\gamma_{el}m_{n}m_{p} ,
\end{equation}
where the momentum of the proton and neutron have been evaluated in the inertial frame of the neutron. Using $m_{p} \sim m_{n}$, we get:

\begin{equation}
\label{gel} \gamma_{el} \sim 2.6.
\end{equation}
If the Lorentz factor of the pick-up ring is such that $1 < \Gamma_r < \gamma_{el}$, the first collisions between an incoming neutron and protons from the ring are more likely to be elastic. After a few such collisions, the relative velocity between the neutron and protons in the ring is reduced, so that the relative kinetic energy becomes lower than the pion production threshold (corresponding to $\gamma_{\pi}$), leading to the shut down of inelastic scattering. The effective relative Lorentz factor under which neutron pick-up by inelastic collision is turned off is therefore $\gamma_{el}$.

For the pick-up ring to be decelerated, we should compare the baryon number of the initial pick-up ring with the full baryon loading received through neutron pick-up (given by Eq. \eqref{dNpuref}). To do so, we evaluate the initial mean free path of a neutron before it is picked-up by inelastic scattering with a proton in the jet.  
Here we have $\Gamma_r = \Gamma_j > \gamma_{el}$, so that an inwardly drifting neutron is most likely to be picked-up from its first collision with a proton from the jet. 

Once a neutron moving sub-relativistically in the transverse direction has crossed the jet boundary it is exposed to an orthogonal, highly relativistic radial flow of baryons. In the laboratory frame (essentially the neutron rest-frame) the corresponding average mean free path is:

\begin{equation}
\label{MFP}\lambda_{inel}(r) = \frac{\beta_{n,in}(r)}{n_{p,j}(r)\sigma_{inel}},
\end{equation}
where $\beta_{n,in}(r)$ is the transverse average inward velocity of a neutron and $n_{p,j}(r)$ is the proton density of the jet near the boundary with the wind at radius $r$ (in the lab frame). The average inward velocity is found by considering an isotropic thermal emission of neutrons from the wind at the border with the jet:

\begin{align}
\label{betanin} \beta_{n,in}(r) &= \int_{0}^{\pi/2} \mathrm{d}\theta \cos(\theta) \sin(\theta) \frac{v_{th}}{c} = \frac{1}{2} {v_{th}/c} \\
& \sim 0.017 \; y^{-1/8} \; \dot{M}_{32}^{1/8} r_{0,7}^{-1/8} \left(\frac{\xi}{1.5}\right)^{1/8} \left(\frac{\beta_{w}}{0.1}\right)^{-1/4} \dot{Q}_{0,19}^{1/8} ,
\end{align}
where the last equality is derived from Eq. \eqref{TwQ}. 

We assumed that the jet already has the total baryon loading necessary for it to be matter dominated with Lorentz factor $\Gamma_j$, and that it is homogeneous, so that the initial jet baryon density in the lab frame is: 

\begin{equation}
\label{defnp0}n_{j,0} \equiv \frac{\delta N_{j}}{\pi \theta_{0}^2 r_{0}^3 \delta y} 
=\frac{L_{j,iso}}{4\pi r_0^2 c \Gamma_j}
\sim 2 \times 10^{27}\; \mathrm{cm}^{-3} \;\; L_{j,iso,52} \Gamma_{j,2}^{-1} r_{0,7}^{-2}.
\end{equation}
With Eq. \eqref{MFP}, this gives the initial inelastic mean free path:

\begin{align}
\label{MFP0}\lambda_{inel,0} &\sim 3\times10^{-4} \; \mathrm{cm} \\
 \nonumber &\hphantom{=}L_{j,iso,52}^{-1} \Gamma_{j,2} \dot{M}_{32}^{1/8} r_{0,7}^{15/8} \left(\frac{\xi}{1.5}\right)^{1/8} \left(\frac{\beta_{w}}{0.1}\right)^{-1/4} \dot{Q}_{0,19}^{1/8} .
\end{align}  
The number of baryons initially in this ring of a shell with radial thickness $\delta y$ of the jet is $\delta N_{j,0} = 2\pi \theta_{0} r_{0}^2 \lambda_{inel,0} n_{j,0} \delta y$:

\begin{equation}
\delta N_{j,0} \sim 4\times10^{37} \delta y \;\; \dot{M}_{32}^{1/8} \theta_{0,-1} r_{0,7}^{2} \left(\frac{\xi}{1.5}\right)^{1/8} \left(\frac{\beta_{w}}{0.1}\right)^{-1/4} \dot{Q}_{0,19}^{1/8} ,
\end{equation}
which is 7 orders of magnitude smaller than the estimated baryon loading from neutron pick-up in Eq. \eqref{dNpuref}. The initial pick-up ring will therefore slow down to $\Gamma_r < \gamma_{el}$ very fast (within $y - 1 \ll 1$). Consequently, the dominant interaction between incoming neutrons and protons from the initial pick-up ring will become elastic scattering: the neutrons are not effectively converted into protons via inelastic scatterings. Because of this effect, we expect the location of the new pick-up ring to steadily move towards the jet axis. 

\subsubsection*{pick-up ring size}

We are now in a position to estimate the size of the pick-up ring at some radius $y$ by equating the diffusion time across the ring with the dynamical time of the jet:

\begin{equation}
\label{tdyn} t_{dyn}(y) = y\frac{r_0}{c} \sim 3 \times 10^{-4} \; \mathrm{s} \;\;\; y \, r_{0,7} \, .
\end{equation} 
We denote by $\tau_{d}$ the typical time necessary for a neutron picked up at radius $y$ to diffuse to a distance $\lambda_{pu}(y)$ from its original position after elastic scatterings in the pick-up region. After a few elastic collisions, the neutron scatterings become isotropic in the frame of the ring with bulk Lorentz factor $\Gamma_r$. The value of $\tau_{d}$ is given by:

\begin{equation}
\label{Td}\tau_{d} = \frac{\lambda_{pu}^2}{\lambda_{el,r}v_{th,r}} \, ,
\end{equation}  
where the ring is now denoted by $r$ and $\lambda_{el,r}$ is the elastic scattering mean free path in the pick-up ring. Note that because we are interested in a displacement transverse to the radial direction, $\lambda_{pu}$ is the same in the ring inertial frame and in the central engine frame. The elastic mean free path is given by:

\begin{equation}
\label{lambels}\lambda_{el,r}(y)= \frac{1}{n_{b,r}\sigma_{el}} \, ,
\end{equation}
with $\sigma_{el}$ the elastic scattering cross-section and $n_{b,r}$ the baryon density in the ring frame.  

The number of picked-up neutrons necessary for the Lorentz factor of the pick-up ring to decrease down to $\Gamma_r$ can be estimated by using the conservation of energy and momentum. We model the neutron pick-up by a collision between the $\delta N_{j,r}$ baryons initially in the ring and the $\delta N_{pu,r}$ picked-up (non-relativistic) neutrons. The conservation of energy and momentum then gives (see for example \citep{Piran1999}, section 8.1.1):

\begin{equation}
\delta N_{j,r} \Gamma_j + \delta N_{pu,r} = (\delta N_{j,r} + \delta N_{pu,r} + \delta e_{int}/c^2) \Gamma_r\, ,
\end{equation}

\begin{equation}
\delta N_{j,r} \sqrt{\Gamma_j^2 - 1} = (\delta N_{j,r} + \delta N_{pu,r} + \delta e_{int}/c^2) \sqrt{\Gamma_r^2 - 1}\, ,
\end{equation}
where $\delta e_{int}$ is the internal energy per unit mass generated in the collision (in the rest frame of the merged mass). We obtain the following relations:

\begin{equation}
\label{dNpus}\delta N_{pu,r} = \Gamma_j \left(\sqrt{\frac{1 - \Gamma_j^{-2}}{1-\Gamma_r^{-2}}} - 1 \right) \delta N_{j,r}\, ,
\end{equation}

\begin{equation}
\label{eint}\delta e_{int}/c^2 =  \left( \frac{\Gamma_j}{\Gamma_r}-1 + \left(\frac{\Gamma_j}{\Gamma_r}-\Gamma_j\right) \sqrt{\frac{1 - \Gamma_j^{-2}}{1-\Gamma_r^{-2}}} \right) \delta N_{j,r}\, , 
\end{equation}
so that the elastic mean free path can be written:

\begin{equation}
\label{lels}\lambda_{el,r}(y) = \Gamma_r\left(\left(1+\Gamma_j\left(\sqrt{\frac{1 - \Gamma_j^{-2}}{1-\Gamma_r^{-2}}} - 1 \right)\right)n_{j,0} \; y^{-2} \sigma_{el}\right)^{-1} \, . 
\end{equation}
\indent The last element we need is the thermal velocity in the inertial frame of the pick-up ring. We expect the pick-up ring to expand very fast against the wind (compared with the dynamical time) until it reaches pressure equilibrium with the wind. This means that lateral expansion happens at fixed $y$. Because the relativistic velocity of the ring is radial, the lateral pressure is the same in the wind frame and in the ring inertial frame, which implies that the ring should reach thermal equilibrium with the wind. $v_{th,r}$ can then be taken to be the thermal velocity in the wind - that is $2\beta_{n,in}c$ in Eq. \eqref{betanin}:  

\begin{equation}
\frac{v_{th,r}(y)}{c} \sim 0.034 \; y^{-1/8} \; \dot{M}_{32}^{1/8} r_{0,7}^{-1/8} \left(\frac{\xi}{1.5}\right)^{1/8} \left(\frac{\beta_{w}}{0.1}\right)^{-1/4} \dot{Q}_{0,19}^{1/8} \, . 
\end{equation}
Note that this lateral expansion also has an impact on the baryon density in the ring $n_{b,r}$, that decreases inversely proportionally to the expansion area, and therefore on the mean free path $\lambda_{el,r}$ in Eq. \eqref{lambels} that increases like the expansion area. Because the angular size of the pick-up ring is much smaller than $\theta_0$, $\lambda_{el,r}$ is actually proportional to the expansion ratio $f_{exp}$ of the initial pick-up ring size. Then the square of the diffusion distance $\lambda_{pu}^2$ increases proportionally to $f_{exp}$ at fixed $\tau_d = t_{dyn}$ from Eq. \eqref{Td}. Therefore the equivalent pick-up ring size before lateral expansion decreases like $f_{exp}^{-1/2}$ as the ring expands. We will refer to the equivalent size before lateral expansion as ``initial size". It is more convenient to use this initial size when discussing the jet's structure, so that unless explicitly specified, we deal only with such quantities. Here, evaluating $\tau_d$ with $f_{exp} = 1$ gives an upper bound on the pick-up ring initial size:

\begin{align}
\frac{\theta_{pu}(y)}{\theta_0} &\equiv \frac{\lambda_{pu}}{y\theta_0 r_0} \\
\nonumber &\lesssim  4\times 10^{-6} \,\, y^{3/8} \\
 \nonumber &\hphantom{=} r_{0,7}^{3/8} \theta_{0,-1}^{-1} L_{j,iso,52}^{-1/2} \dot{M}_{32}^{1/8} \left(\frac{\xi}{1.5}\right)^{1/8} \left(\frac{\beta_{w}}{0.1}\right)^{-1/4} \dot{Q}_{0,19}^{1/8}  , 
\end{align} 
where we used the fact that the Lorentz factor of the pick-up ring should be smaller than $\gamma_{el}$. From $y_{r-pro}$, radioactive heating ceases in the wind, so that the temperature drops, reducing the diffusion in both the wind and the pick-up ring. $y_{r-pro}$ is therefore the relevant radius to evaluate the angular size of the pick-up ring. We find

\begin{align}
\label{ubtc}\frac{\theta_{pu}(y_{r-pro})}{\theta_0}  &\lesssim 4\times  10^{-5} \,\, \theta_{0,-1}^{-1} L_{j,iso,52}^{-1/2} \dot{M}_{32}^{1/8}\\ \nonumber &\hphantom{=} \left(\frac{\xi}{1.5}\right)^{1/8} \left(\frac{\beta_{w}}{0.1}\right)^{1/8} \dot{Q}_{0,19}^{1/8} \left(\frac{\Delta t_{r-pro}}{1\, s}\right)^{3/8} \, .
\end{align}
With this size, Eq. \eqref{dNpus} tells us that the main pick-up ring is not relativistic. More precisely, we find:

\begin{equation}
\label{ubgs}1-\Gamma_r^{-1} \lesssim 0.1 \,\, \left(\frac{1-Y_e}{2\sqrt{Y_e}}\right)^{-2}\dot{M}_{32}^{-1} \left(\frac{\xi}{1.5}\right)^{-1} L_{j,iso,52} \, . 
\end{equation}
With the current assumption that neutrons become protons but protons do not become neutrons, the pick-up ring would not be relevant to sGRB 170817A and its afterglow, because most of the pick-up ring is non-relativistic. Purely elastic diffusion fails to explain the jet structure observed in GW170817 because the pick-up ring becomes saturated with neutrons too fast, impeding them to go further towards the core of the jet. 
However, in the next subsection we include a process neglected up to now, that is the inelastic conversion of protons into neutrons:
\begin{equation}
\label{npinel2} n + p \to n + n + \pi^{+} \, .
\end{equation}
This makes it possible for a conversion-diffusion process to occur and drive baryons further towards the core of the jet.

\subsection{The conversion-diffusion model}

\label{sec:convdiff}

\begin{figure}
\begin{center}
\includegraphics[scale = 0.5]{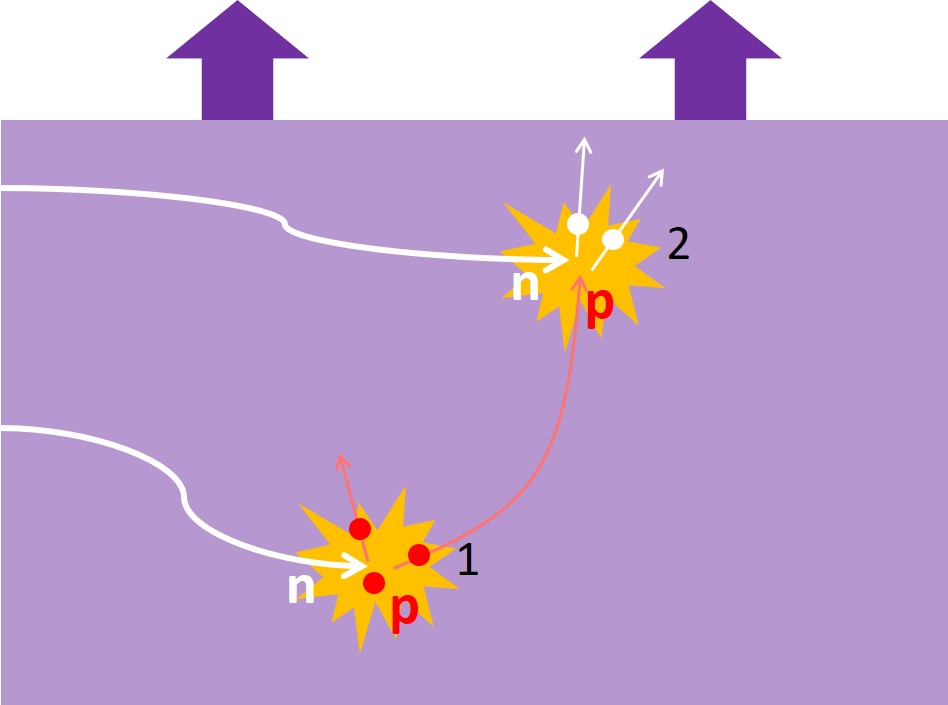}
\end{center}
\caption{Schematic representation of the conversion-diffusion process. The purple region corresponds to the pick-up ring, with the wind at the left, the
thick purple arrows denoting the bulk velocity of the jet.  We follow here the trajectory of a specific incoming neutron that has leaked from the wind. At 1, it collides inelastically for the first time with a proton from the jet, and is converted into a proton, which then follows the local fluid line of the jet that goes together with a magnetic line. In this state it is impossible for the proton to keep drifting orthogonally to the radial direction. At 2, it collides inelastically with a neutron and is converted back to a neutron. It can then keep drifting towards the core of the jet until it collides again with a proton.}
\label{fig:convdiff}
\end{figure}

The conversion mechanism has already been studied in other contexts \citep{DerishevAharonianKocharovskyEtAl2003, KashiyamaMuraseMeszaros2013} but here we apply it for the first time explicitly in the context of the diffusion process. The principle is summarized in Fig. \ref{fig:convdiff}. The main difference with the elastic scattering diffusion model is that protons can be converted to neutrons again, and thus can diffuse further towards the core of the jet. Of all the inelastic collisions that happen between protons from the pick-up ring and incoming neutrons, there is indeed a non-negligible fraction that results in the conversion of the proton into a neutron. The latter is able to move transversely to the radial direction, so that protons from the pick-up ring are (in the form of neutrons) effectively able to continue diffusing. This makes the diffusion process more efficient, especially as far as the relativistic part of the pick-up ring is concerned. Note that this mechanism further implies that the pick-up ring also fills up with neutrons, which are also susceptible of scattering with the incoming neutrons elastically or inelastically, via 
\begin{equation}
\label{nninel} n + n \to n + p + \pi^{-} \,\, , \,\, n + n \to n + n + \pi^0\, .
\end{equation}

For the conversion-diffusion process to be effective, the relative Lorentz factor $\Gamma_{rel}$ between cylindrical radii $x$ and $x + \lambda_{inel}(x,y)$ - where $\lambda_{inel}(x,y)$ is the inelastic mean free path of neutrons emitted from $x$ towards the jet's axis - should be sufficiently large for inelastic collisions to occur within the ring. In particular it should be such that $\Gamma_{rel} > \gamma_{\pi^-}$, where $\gamma_{\pi^-}$ is defined in Eq. \eqref{gp}. Note that the Lorentz factor for the inward velocity of neutrons emitted after a scattering event is also of the order of $\Gamma_{rel}$. Therefore, in order for the inelastic scattering mean free path to be smaller than the elastic one, we should actually have the stronger constraint $\Gamma_{rel} \gtrsim \gamma_{el}$, where $\gamma_{el}$ is defined in Eq. \eqref{gel}. A detailed analysis (not presented here) indicates that the conditions in the conversion-diffusion pick-up ring are such that  $\Gamma_{rel}$ does not depend strongly on $x$, and remains close to $\gamma_{el}$. In what follows, we assume for simplicity that $\Gamma_{rel}$ is approximately \emph{constant} equal to $\gamma_{el}$.\footnote{Numerical results indicate that higher values of $\Gamma_{rel}$ seem to imply an enhanced diffusion. Considering $\Gamma_{rel} = \gamma_{el}$ may therefore be a conservative choice.} Then, the mean free path for conversion-diffusion $\lambda_{inel}$ imposes a typical variation scale for the Lorentz factor as a function of the cylindrical radius from the axis of the jet: 
\begin{equation}
\label{gammarelcons}\frac{\Gamma(x) - \Gamma'(x)\lambda_{inel}}{\Gamma(x)} \sim \Gamma_{rel} + \sqrt{\Gamma_{rel}^2 - 1}  \, , 
\end{equation}
where we consider the region of the ring where $\Gamma(x)^2 \gg 1$. We get:

\begin{equation}
\label{Gld}\left|\frac{\Gamma(x)}{\Gamma'(x)}\right| \sim \frac{1}{\Gamma_{rel} + \sqrt{\Gamma_{rel}^2 - 1}-1}\lambda_{inel}(x,y)\, . 
\end{equation}
  
Interestingly, the mean free path itself depends on the Lorentz factor as (in annalogy with Eq. \eqref{lels}):

\begin{eqnarray}
\nonumber \lambda_{inel}(x,y) &\sim& \frac{\Gamma(x)}{n_{j,0}y^{-2} \sigma_{inel}} \left(1+\frac{\left(\Gamma_{rel} + \sqrt{\Gamma_{rel}^2 - 1}\right)\Gamma_j}{\Gamma(x)} \right. \\ 
\label{ldxygen} &\hphantom{=}& \left.\frac{\sqrt{\frac{1 - \Gamma_j^{-2}}{1-\Gamma(x)^{-2}}} - 1 }{1-\sqrt{\frac{1 - \left(\Gamma_{rel} + \sqrt{\Gamma_{rel}^2 - 1}\right)^2\Gamma(x)^{-2}}{1-\Gamma(x)^{-2}}}} \right)^{-1}\, , 
\end{eqnarray}
where we used the generalization of Eq. \eqref{dNpus}-\eqref{eint} in the case where the incoming neutrons at $x$ come from the neighboring ring with Lorentz factor $\Gamma(x+\lambda_{inel})$,

\begin{equation}
\label{dNpusbis}\delta N_{in,r} = \frac{\Gamma_j}{\Gamma(x+\lambda_{inel})}\, \frac{\sqrt{\frac{1 - \Gamma_j^{-2}}{1-\Gamma(x)^{-2}}} - 1 }{1-\sqrt{\frac{1 - \Gamma(x+\lambda_{inel})^{-2}}{1-\Gamma(x)^{-2}}} }\,\, \delta N_{j,r} \, , 
\end{equation}

\begin{align}
\label{eintbis}\frac{\delta e_{int}}{c^2} &=  \left( \frac{\Gamma_j}{\Gamma(x)}-1 \right. \\
\nonumber &\hphantom{=} \left. + \left(\frac{\Gamma_j}{\Gamma(x)}-\frac{\Gamma_j}{\Gamma(x+\lambda_{inel})}\right) \frac{\sqrt{\frac{1 - \Gamma_j^{-2}}{1-\Gamma(x)^{-2}}} - 1 }{1-\sqrt{\frac{1 - \Gamma(x+\lambda_{inel})^{-2}}{1-\Gamma(x)^{-2}}} } \right) \delta N_{j,r} \, , 
\end{align}
together with  Eq. \eqref{Gld} to find:
\begin{equation}
\label{linap}\Gamma(x-\lambda_{inel}) \sim \Gamma(x) - \lambda_{inel} \Gamma'(x) \sim \left(\Gamma_{rel} + \sqrt{\Gamma_{rel}^2 - 1}\right) \Gamma(x)\\ \, ,
\end{equation}
so that $\Gamma(x+\lambda_{inel}) \sim \Gamma(x)/\left(\Gamma_{rel} + \sqrt{\Gamma_{rel}^2 - 1}\right)$\footnote{Although we are not in a regime for which the linear approximation \eqref{linap} is very accurate $\left(\text{as} \left|\frac{\Gamma(x)}{\Gamma'(x)}\right| \sim \frac{1}{\Gamma_{rel} + \sqrt{\Gamma_{rel}^2 - 1}-1}\lambda_{inel}(x,y)\right)$, one can check that locally the resulting error on $\Gamma(x+\lambda_{inel})$ is less than 30\% in the solution represented in Fig. \ref{gi}.}. 
Note that, unlike Eq. \eqref{dNpus}, the quantity $\delta N_{in,r}$ in Eq. \eqref{dNpusbis} is not the number of baryons picked-up in the sub-ring at $x$, but rather the number of incoming neutrons that collide with the baryons in this sub-ring. Because of the reactions that have neutrons as products, the latter number is expected to be somewhat larger (by a factor $\lesssim 3$, as can be estimated in the framework developed at the end of Appendix \ref{sec:checks}) than the pick-up number $\delta N_{pu,r}$ that sets the local baryon density. For simplicity, this difference is ignored in Eq. \eqref{ldxygen} .

\indent For $\Gamma_{rel} = \gamma_{el}$, $\Gamma_j \gg \Gamma(x)$ and $\Gamma(x)^2 \gg 1$, Eq. \eqref{ldxygen} reduces to:

\begin{equation}
\label{ldxy}\lambda_{inel}(x,y) \sim \frac{1}{\frac{5\Gamma_j}{24\Gamma(x)^2} n_{j,0}y^{-2} \sigma_{inel}} \, , 
\end{equation}
which gives the typical profile for the Lorentz factor in the pick-up ring for this type of diffusion with Eq. \eqref{Gld}:

\begin{equation}
\label{Gfit}\Gamma(x,y) \sim  \left( 1 + \frac{5}{3} \Gamma_j n_{j,0} y^{-2} \sigma_{inel} (r\theta_0 - x)\right)^{1/2} \, . 
\end{equation}

\section{Jet structure and implications for GRB 170817A}

\begin{figure}
\begin{center}
\includegraphics[scale = 0.5]{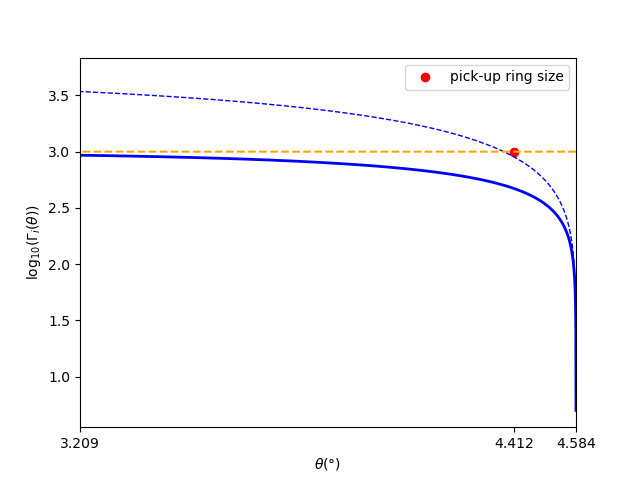}
\caption{Lorentz factor in the pick-up ring as a function of the polar angle. We choose $y = y_{r-pro}$ and $\Gamma_j = 1000$, $\Delta t_{r-pro} = 1s$ and $\theta_0 = 0.08\, \mathrm{rad}$. The dashed blue line represents the solution of Eq. \eqref{Gldgen} and the full blue line the solution obtained from imposing that the Lorentz factor go to $\Gamma_j$ in the jet core (we do this concretely by removing the leftmost 1 in the right hand side of Eq. \eqref{Gldgen}). The exact solution should be somewhere between these plotted lines. As discussed in the main text, though, either choice yields the same result for the off-axis structure at the relevant angles.} 
\label{gi}
\end{center}
\end{figure}

We now derive the resulting Lorentz factor and isotropic equivalent energy structure after the jet breaks out of the ejecta and the pick-up ring has expanded, in order to compare it with GRB 170817A. Note that equation \eqref{Gfit} is only valid for $\Gamma_j \gg \Gamma(x)$, that is sufficiently close to the boundary with the wind. To obtain the exact solution to equation \eqref{Gld}, one needs to solve numerically the following equation:

\begin{equation}
\label{Gldgen}\Gamma'(\theta) = - n_{j,0} r_0 \sigma_{inel} y^{-1} \left(1+ \frac{5\Gamma_j}{\Gamma(x)}\, \frac{\sqrt{\frac{1 - \Gamma_j^{-2}}{1-\Gamma(x)^{-2}}} - 1 }{1-\sqrt{\frac{1 - 25\Gamma(x)^{-2}}{1-\Gamma(x)^{-2}}} }\right) \, , 
\end{equation}
where we changed variables from the cylindrical radius $x$ to the polar angle $\theta$. Solving this equation gives the Lorentz factor structure of the pick-up ring before the post-breakout expansion (that is also before the first phase of lateral expansion against the wind due to the pressure of the ring discussed in Sec.~\ref{eldif} and Appendix \ref{B}). The result is represented in Fig.~\ref{gi}. Note that the Lorentz factor goes up to $5\Gamma_j$ (where Eq. \eqref{Gldgen} vanishes) instead of $\Gamma_j$ in the jet core, as shown by the blue dashed line. This is an artifact of the fact that Eq. \eqref{Gldgen} gives the local behavior of $\Gamma(x)$ which is compatible with the conversion-diffusion process, but does not take into account boundary effects (which are subject to uncertain assumptions). Nevertheless, the boundary effects do not modify the solution for $\Gamma(x) < 100$, and since the main emitting region for the signal observed at $\theta_v$ should have Lorentz factor $\Gamma \sim 10$, the blue dashed solution shown in Fig. \ref{gi} should be appropriate for describing the relevant off-axis structure. A corresponding solution enforcing a correct initial value of $\Gamma_j$ is shown as a full blue line in the same figure. Note also that the former choice is an upper bound on the real solution and is therefore conservative. 

\begin{figure}
\begin{center}
\includegraphics[scale = 0.5]{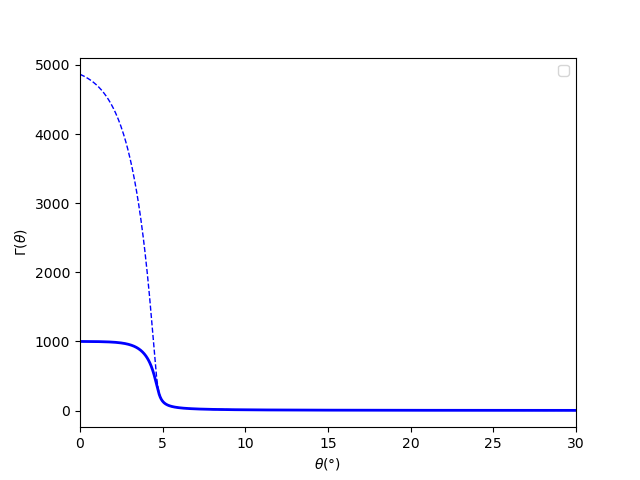}
\caption{Lorentz factor structure after expansion of the pick-up ring, for $\Gamma_j = 1000$, $\Delta t_{r-pro} = 1s$ and $\theta_0 = 0.08\, \mathrm{rad}$. The dashed blue line represents the solution from Eq. \eqref{Gldgen} and the full blue line the solution obtained from imposing that the Lorentz factor go to $\Gamma_j$ in the jet core (we do this concretely by removing the leftmost 1 in the right hand side of Eq. \eqref{Gldgen}). It is pretty clear on this figure that either choice yield the same result for the Lorentz factor structure at the relevant angles with $\Gamma \sim 10$.}
\label{gf}
\end{center}
\end{figure}

After the post-breakout expansion, each ring with a Lorentz factor $\Gamma$ expands from the initial angle $\theta_r(\Gamma)$ to an angle $\theta_r + 1/\Gamma$. The Lorentz factor angular dependence right after expansion is therefore given by:

\begin{equation}
\label{Gammaf}\Gamma(\theta) = \frac{1}{\theta - \theta_r(\Gamma)} \, . 
\end{equation}
Then thermal energy is converted into kinetic energy in the expanded ring, so that $\Gamma(\theta)$ is accelerated to $\Gamma_f(\theta)$. According to the estimate in Eq. \eqref{usem}, the thermal energy is comparable to the rest mass energy in the pick-up ring. So $\Gamma_f \sim \Gamma$, and Eq. \eqref{Gammaf} gives the final Lorentz factor distribution. The resulting structure for the Lorentz factor after expansion is represented as the blue dashed line in Fig. \ref{gf}. Again, as noted before, the dashed blue line Lorentz factor goes up to $5\Gamma_j$ instead of $\Gamma_j$ in the jet core, due to the fact that the final Lorentz factor structure is derived from the initial structure plotted in Fig. \ref{gi}. As we discussed there, this inconsistency near the jet axis does not influence the part of the structure that is relevant to the off-axis GRB, being a conservative choice. A corrected solution is shown with the full blue line.

For a single component of the pick-up ring initially at angle $\theta_r$ from the jet's axis, the isotropic equivalent energy changes after expansion like: $E_{j,iso} \to |\theta_r \mathrm{d}\theta_r / ((\theta_r + \Gamma^{-1}) \mathrm{d} \theta_f)| E_{j,iso}$. With $\theta_r(\Gamma)$ given by equation \eqref{Gfit}, we deduce:

\begin{equation}
\label{Eisof}E_{iso}(\theta) \sim \frac{6\Gamma(\theta)^4}{5\Gamma_j} \frac{\theta_0 y_{r-pro}}{n_{j,0}\sigma_{inel} r_0} E_{j,iso} \, . 
\end{equation}
Note that this approximation is valid for large angles\footnote{From Eq. \eqref{Gldgen} the derived structure is actually defined only for $\Gamma \gtrsim 5$. For larger angles we propose to extrapolate the structure, assuming Eq. \eqref{Eisof} still holds there. This choice is itself a part of our structure model and may in principle be replaced by a faster transition to a steep decline. Numerical simulations would be a way of getting insight on the right choice for the large angle structure.}. The general solution is obtained from Eq. \eqref{Gldgen}.

\subsubsection*{Off-axis emission}

We follow \citet{IokaNakamura2019} to derive the observed isotropic equivalent energy $E_{off}$ at viewing angle $\theta_v$ after taking into account the off-axis emission outside the beaming angle. Equation (11) in their paper gives a relation with $E_{iso}(\theta)$:

\begin{equation}
\label{Eoffaxis}E_{off}(\theta_v) = \int_0^{\pi/2} \frac{\sin(\theta)\mathrm{d}\theta}{2} E_{iso}(\theta) \mathcal{B}(\theta,\theta_v) \, , 
\end{equation}
where the beaming parameter is defined by:

\begin{align}
\label{defB}\mathcal{B}(\theta,\theta_v) & \equiv \int_{-\pi}^{\pi}\frac{\mathrm{d}\phi}{2\pi} \frac{1}{\Gamma_f(\theta)^4 (1-\beta(\theta) \cos\theta_{\Delta})^3} \, ,  \\
& = \frac{1}{2\Gamma_f(\theta)^4} \frac{2(1-\beta(\theta)\cos\theta\cos\theta_v)^2 + (\beta(\theta)\sin\theta\sin\theta_v)^2}{(1-\beta(\theta)\cos(\theta_v+\theta))^{5/2} (1-\beta(\theta)\cos(\theta_v-\theta))^{5/2}} \, , 
\end{align}
where $\Gamma_f$ is defined in Eq. (\ref{Gammaf}) and $\theta_{\Delta}$ is the angle between the point at $(\theta,\phi)$ and the line of sight at $(\theta_v,0)$, which obeys:

\begin{equation}
\label{eqthetaD}\cos\theta_{\Delta} = \sin\theta_v\sin\theta\cos\phi + \cos\theta\cos\theta_v  \, . 
\end{equation}
Note that, due to compactness considerations \citep{MatsumotoNakarPiran2018, MatsumotoNakarPiran2019}, the region of the structured jet that contributes to the observed signal of GRB 170817A may be smaller than in Eq. \eqref{Eoffaxis}. Because we check that the region near the jet core is negligible to the off-axis emission, the off-axis emission computed here should be at worst slightly overestimated.

\begin{figure}
\begin{center}
\includegraphics[width=8cm]{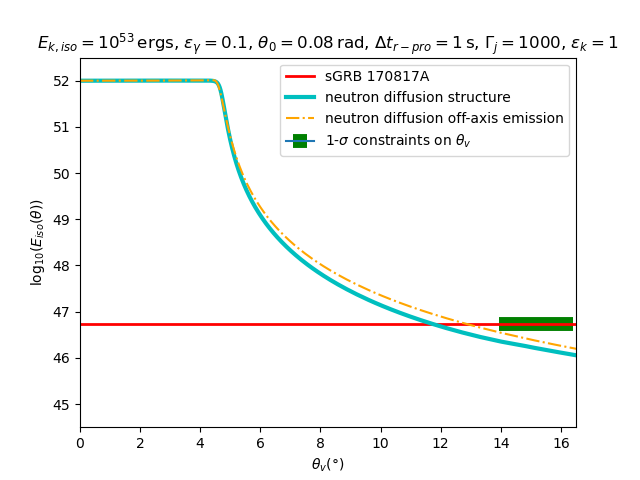}
\caption{Isotropic equivalent energy structure as a result of neutron conversion-diffusion from the wind and the lateral expansion after the jet breakout from the ejecta (blue line). Values for the parameters are written above, where $\epsilon_k$ is the fraction of the jet's energy that goes into the kinetic energy of baryons. The red line indicates the observed isotropic equivalent energy for sGRB 170817A and the orange dotted line is the integrated off-axis emission at the observation angle. The 1-$\sigma$ constraints on the observation angle $\theta_v$ are represented in green.}
\label{se}
\end{center}
\end{figure}

Fig. \ref{se} shows the result for the gamma-ray isotropic equivalent energy structure as a function of the viewing angle for $\Gamma_j = 1000$. We assumed a uniform gamma-ray efficiency $\epsilon_{\gamma} \sim 0.1$. Constraints on the viewing angle are taken from  \citet{MooleyDellerGottliebEtAl2018}.

It is apparent from this figure that taking only neutron diffusion cannot explain the observed GRB 170817A for a purely baryonic top-hat jet. In this case, the pick-up ring is too thin to affect the emission of sGRB 170817A.

\subsubsection*{Neutron diffusion in a structured jet}

In contrast to the top-hat case, the neutron diffusion is efficient if the jet core is structured and its density is low at the boundary with the wind. Because of its compatibility with the slow rise of the afterglow light curves associated with GW170817, we consider a Gaussian structure for the jet core isotropic energy. Note that the afterglow observations constrain the jet structure only near the jet core, not in the outer part around the line-of-sight \citep{Takahashi+19}. The bulk Lorentz factor structure is assumed to be described by a power-law. The structure is therefore parametrized as follows:
\begin{align}
E_{k,iso,core} &= E_0 \exp(-\theta^2/2\theta_{0}^2), \\
\Gamma_{core} &= \frac{\Gamma_{max}}{1+(\theta/\theta_{0})^\lambda}\, .
\end{align}
We further define $\theta_j$ as the angular size of the jet core. We should note that a weak outer jet is stripped during the propagation through the ejecta before the breakout because the weak jet head is much slower than the main jet head and the cocoon pressure is also high. However after the jet breakout, the cocoon pressure decreases drastically and the weak jet can keep its structure. 

\begin{figure}
\begin{center} 
\includegraphics[width=8cm]{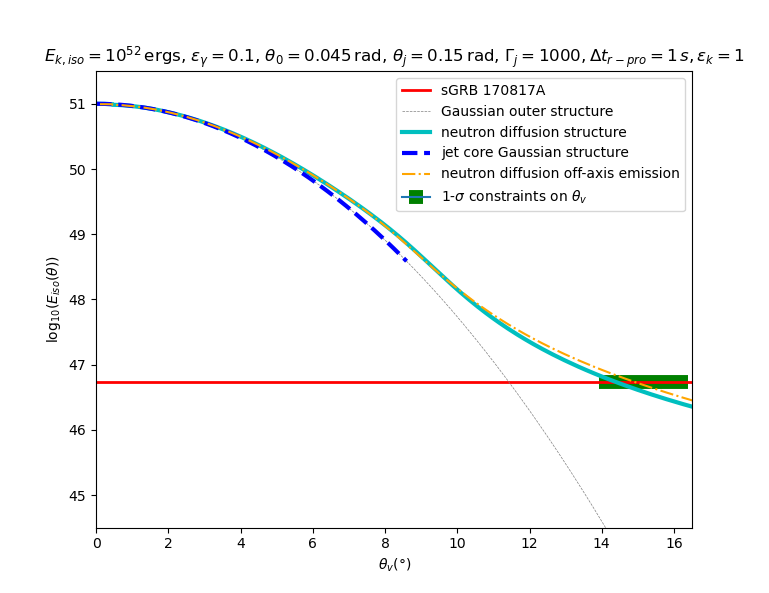}
\caption{Isotropic equivalent energy structure for a Gaussian jet core with Lorentz factor $\Gamma_{max} = 1000$ and angular size $\theta_j = 0.15\, \mathrm{rad}$ (full light cyan line). Values for the parameters are written above. The dotted orange line corresponds to the integrated off-axis emission. We also plot the observed isotropic equivalent energy for sGRB 170817A (red line), the jet-core Gaussian structure (dashed blue line), the corresponding full Gaussian structure for comparison with neutron pick-up (dashed grey line) and the 1-$\sigma$ constraints on the observation angle $\theta_v$ in green.}
\label{Esl}
\end{center}
\end{figure}

We then consider the outer structure induced by neutron diffusion on such a Gaussian jet core. In this case, because neutron diffusion is only sensitive to the density and Lorentz factor very close to the boundary with the wind (the precise condition is that the pick-up ring size $\theta_{pu}$ is such that $\Gamma_{core}'(\theta_j)\theta_{pu} \ll \Gamma_{core}(\theta_j)$ ), the parameters $\Gamma_j$ and $E_{j,iso}$ in Eq. \eqref{Eisof} correspond to the Lorentz factor and the isotropic equivalent energy of the Gaussian jet core at the boundary angle $\theta_j$, respectively. Fig. \ref{Esl} shows the corresponding isotropic equivalent energy profile for a core with $\theta_0 = 0.045\, \mathrm{rad}$, $E_0 = 10^{52}\, \mathrm{erg}$, $\Gamma_{max}=1000$, $\lambda = 3$ and $\theta_j = 0.15\, \mathrm{rad}$. Note that near the boundary with the wind, we then have $\Gamma_j \sim 30$ and $E_{j,iso} \sim 10^{48} \, \mathrm{erg}$. 

We see that with this structure for the jet core, the off-axis emission from the neutron-diffusion-induced structure dominates the Gaussian structure and is consistent with GRB 170817A, at the viewing angle consistent with the observations within 1 sigma error. This means that if the jet core has a structure, neutron diffusion could be essential for explaining the sGRB170817A observation. \\
\indent Furthermore, the power-law behavior for the neutron diffusion structure in Eq. \eqref{Eisof} $E_{iso,f} \sim \theta^{-4}$ (thick light blue line in Fig.~ \ref{Esl}) is shallower than any Gaussian jet, and also than the best power-law fit of \cite{GhirlandaSalafiaParagiEtAl2018}, where they find a power $s_1 \sim 5.5_{-1.4}^{+1.3}$. This means that neutron pick-up is not only a good candidate scenario for the jet structure of sGRB170817A, but should also be the dominant contribution at larger angles, even if the jet structure near the axis is due to some other phenomenon.

\subsubsection*{Magnetized jet}

\begin{figure}
\begin{center}
\includegraphics[width=8cm]{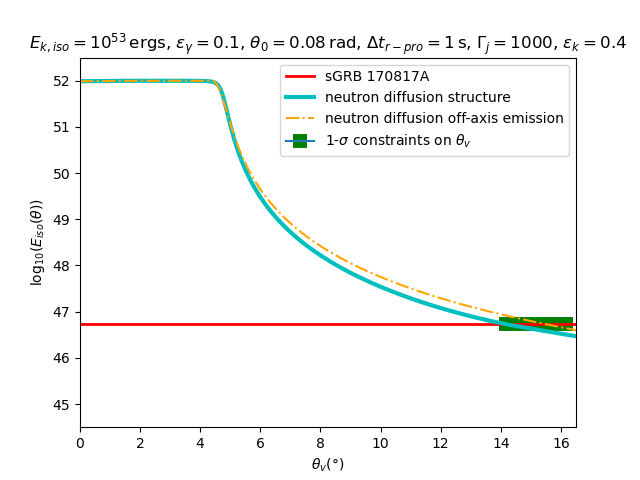}
\caption{Isotropic equivalent energy structure as a result of neutron diffusion from the wind and the lateral expansion after the jet breakout from the ejecta for a magnetized jet with $\epsilon_k = 0.4$ (blue line). Values for the parameters are written above. The red line indicates the observed isotropic equivalent energy for sGRB 170817A and the orange dotted line the integrated off-axis emission at the observation angle. The 1-$\sigma$ constraints on the observation angle $\theta_v$ are represented in green.}
\label{Em}
\end{center}
\end{figure}

From Eq. \eqref{Eisof} for a top-hat jet,  we see that the observed isotropic equivalent energy scales proportionally to the inverse of the initial baryon density in the jet $n_{j,0}$ (which corresponds to the fact that the diffusion length in the jet $\lambda_{inel}$ has this scaling). The neutron diffusion model is therefore sensitive to the magnetization of the jet, that is the fraction of the jet energy that goes into the magnetic field - or inversely, the fraction of the jet energy that goes into the kinetic energy of baryons, which we denote $\epsilon_k$. It is then worth investigating the magnetized regime $\epsilon_k < 1$ \citep{MeszarosRees2011}, which is not unnatural if the jet is launched through a Blandford-Znajek mechanism. Fig. \ref{Em} shows the corresponding isotropic equivalent energy profile for $\epsilon_k = 0.4$ and $\Gamma_j = 1000$. 

In this case, the off-axis emission from the jet is consistent with GRB 170817A at a viewing angle consistent with the observations within $1$-$\sigma$ error. It can be checked that the angular separation between the main emitting region and the observation angle is consistent with the constraint imposed by compactness considerations (from \citet{MatsumotoNakarPiran2019}) - typically $0.1 \, \mathrm{rad}$.

\section{Comparison with other jet structures from the literature}

We compare in this section the jet structure resulting from neutron conversion-diffusion to other jet structures considered in the literature. The structure for the isotropic equivalent energy that emerges within our model is expressed at large angles by Eq. \eqref{Eisof}, with $\Gamma(\theta)$ given in Eq. \eqref{Gammaf}, that is a power-law $E_{iso}(\theta) \sim \theta^{-4}$. 

We first note that our jet energy structure is much less shallow than the original universal power-law structure model studied by \citet{Rossi2002}, \citet{Zhang2002} and \citet{Perna2003}. To reproduce the relation between isotropic energy and jet break time observed in GRB afterglows, it was found there that the power $\alpha_{\epsilon}$ for the energy structure had to obey 
\begin{equation}
\label{conpl} 1.5 \lesssim \alpha_{\epsilon} \lesssim 2.2 \, , 
\end{equation}
at the 1-$\sigma$ level. Our model is not power-law near the jet center though, where there is a well defined jet core of definite size $\theta_0$. The power-law behavior appears only for $\theta \gtrsim 2 \theta_0$ and the jet break time is still determined by the properties of the jet core. The observational constraint of Eq. \eqref{conpl} therefore does not apply to the structure induced by neutron conversion-diffusion.

Interestingly, the structure from our model is similar to what was observed in the recent GRMHD simulations of the propagation of hydrodynamical jets in the ejecta of neutron star mergers by \citet{Gottlieb2021}. They found that the structure could be divided into three components: the jet core (flat in their simulation), the jet-cocoon interface (JCI) where the structure is power-law up to about $\theta \sim 17\degree$ and the cocoon where the isotropic energy decays exponentially. Figs. \ref{se} and \ref{Esl} show the same kind of structure, except for the cocoon which is not taken into account by our model. Also, the power-law in the JCI in their simulations of sGRBs is shallower than $\theta^{-4}$, with a power $-3.5 \lesssim \alpha_{\epsilon}\lesssim -3$. In their case, the formation of the JCI is also due to the loading of the jet in baryons at the boundary with the wind, but instead of neutron-conversion diffusion the relevant process at stakes is that of hydrodynamic mixing. The fact that they find a shallower power-law therefore tends to indicate that hydrodynamic mixing may be more efficient than neutron conversion-division to load the jet with baryons and produce a structure. In another set of simulations \citep{Gottlieb2020} they find that the presence of magnetization in the jet - even small - implies a sizable reduction in the mixing with the wind material. Depending on the magnetization of the jet, neutron conversion-diffusion may therefore dominate over hydrodynamic mixing. If the mixing leaves a sufficiently sharp Lorentz factor profile at the boundary between the JCI and the wind, it could also be that a hydrodynamic profile would be followed by an outer neutron conversion-diffusion structure before the transition to the cocoon. Note that a power-law structure $E_{iso}(\theta) \sim \theta^{-3.5}$ similar to that of \citet{Gottlieb2021} has been observed in the 3D GRMHD simulations of \citet{Kath2019}. By contrast, they found the jet structure to be hollow, with a slow core of size $\theta_{\text{core}}\sim 3\degree$, which is also consistent with the observations \citep{Takahashi+19,takahashi2020diverse}. This last feature was reproduced in the 3D GRMHD simulations of  \citet{Nathanail2020, Nathanail2021}.

The simulations of \citet{Lazzati2017} also present interesting results for the jet structure. They studied the propagation of an sGRB jet in an ejecta where the particle density profile is modelled by an exponential decay with growing distance from the central engine, and found the resulting jet structure to be well described by a sum of two exponentials 
\begin{equation}
\label{sesL} \frac{\mathrm{d}E}{\mathrm{d}\Omega} = A e^{-\frac{\theta}{\theta_j}} + B e^{-\frac{\theta}{\theta_c}} \, ,
\end{equation}
and likewise for the Lorentz factor structure. As for \citet{Gottlieb2021}, the structure can be divided into the structured jet (with size $\theta_j$ in Eq. \eqref{sesL}) on the one hand and the cocoon on the other hand (with size $\theta_c$ in Eq. \eqref{sesL}) but there is no clear distinction between the jet core and the interface with the cocoon. The steep decline of the Lorentz factor at the transition between the jet and the cocoon in this structure favours the existence of an outer neutron conversion-diffusion profile as in Fig. \ref{Esl}. 

We finally quote the 2D axially symmetric relativistic hydrodynamic simulations of \citet{Xie2018}, where the radial profile of the density in the ejecta is modelled by a power-law with power $-8$. The structure they observed can also be divided into two components, one being the structured jet and the other the cocoon. In their case the jet energy structure is found to be quasi-Gaussian and expands up to an angle of about $\sim 15\degree$, where the Lorentz factor decays exponentially fast. This configuration is actually close to the particular example we considered when discussing neutron diffusion in a structured jet core (with the resulting neutron converion-diffusion structure plotted in Fig. \ref{Esl}) so we expect it to favour the existence of an outer neutron conversion-diffusion structure at large angles.

\section{Summary and discussion}

In this paper we proposed a model for the structure of the jet observed in GW170817, arguing that this structure may be the consequence of the pick-up of neutrons leaking from the ejecta into the jet. Studying the diffusion of neutrons into the jet, we first showed that the baryon loading is dominated by picked-up neutrons very close to the boundary between the wind and the jet. We argued that considering elastic scattering as the only diffusion process of picked-up neutrons towards the center of the jet fails to explain a jet structure compatible with observations. Then we considered the neutron-proton conversion mechanism and the corresponding conversion-diffusion mechanism (see Fig. \ref{fig:convdiff}), which speeds up the inward neutron diffusion drastically. We finally estimated the observed isotropic equivalent energy of the emitted photons and compared this with the observational data for sGRB 170817A. The conclusion is that neutron conversion-diffusion induces a relatively shallow power-law structure $E_{iso} \sim \theta^{-4}$ (from Eq. \eqref{Eisof}), which always dominates the outer structure for a Gaussian-like jet. We showed that the neutron conversion-diffusion in a non-magnetized top-hat jet is not effective for explaining sGRB 170817A. However the neutron conversion-diffusion can be essential for determining the generic jet outer structure, and in particular sGRB 170817A, if the jet core is structured with a sufficiently weak tail, e.g. a Gaussian, or if the jet is magnetized, because the density of the jet is effectively low at the boundary in these cases. Future detailed numerical simulations would be desirable to substantiate some of the analytical arguments presented here, one of the things to check being the importance of turbulent convection in the transfer of baryons
in comparison to the conversion-diffusion effect.

We emphasize that the neutron conversion-diffusion by itself cannot explain the observed afterglow which puts constraints on the structure of the jet core (if the jet is not magnetically-dominated), but it could be the origin of the structure from which the prompt GRB is emitted. Because the inferred structure is relatively shallow compared with the current models used to discuss the afterglow light curves, the effect of neutron conversion-diffusion discussed here actually dominates the off-axis emission at large angles.

Finally, we point out that the detection of off-axis sGRBs and afterglows in the future should make it possible to test our model. In particular, the observation of the (shallow $E_{iso,f} \propto \theta^{-4}$) outer structure would be a strong argument in favor of this neutron conversion-diffusion model. 
\\

\section*{Acknowledgements}
 
K.~I. acknowledges support from JSPS KAKENHI Nos. 20H01901, 20H01904, 20H00158, 18H01213, 18H01215, 17H06357, 17H06362, 17H06131,
and P.M. acknowledges support from the Pennsylvania State University Eberly Foundation.

\section*{Data Availability}

No new data were generated or analysed in support of this
research.

\bibliographystyle{mnras}

\bibliography{Bib_sGRB}

\label{lastpage}

\appendix

\section*{Appendix}

\section{Jet model and parameter constraints\label{A}}

The jet is assumed to be kinematically dominated, described by the standard fireball model (e.g., \cite{Piran1999,Meszaros2006}) except for the magnetized jet case. We denote by $t_j$ the time after the merger when the jet is launched. We suppose that it is already matter dominated at $r_0$. Also, we assume that the main baryon loading of the jet has already taken place before neutron leakage from the wind and that it is homogeneous for a top-hat jet. Note that for this assumption to be consistent, we will need to check that the total baryon loading from neutron leakage is negligible compared with the total baryon number in the jet.

The afterglow observations put constraints on the isotropic energy of the jet, the opening angle of the jet, and the viewing angle between the line-of-sight and the jet axis \citep{afterglow2018, MooleyDellerGottliebEtAl2018, GhirlandaSalafiaParagiEtAl2018, HotokezakaNakarGottliebEtAl2018, WuMacFadyen2019}. The ranges considered correspond to 1-sigma intervals of confidence.   
\begin{itemize}
\item \underline{$E_{k,iso}$}: the on-axis kinetic isotropic equivalent energy. It typically lies in the range $E_{k,iso} \sim 10^{52}-10^{53}$ ergs . Note that it does not give directly the isotropic equivalent energy associated with the gamma-ray burst emission. The latter is obtained after multiplication by the gamma-ray conversion efficiency $\epsilon_{\gamma}$, that is generally estimated to be of the order of $0.1$ 

\item \underline{$\theta_0$}: the jet's opening angle.  

\item \underline{$\theta_v$}: the observation angle, measured from the axis of the jet. This parameter is not a parameter of the jet model but it is crucial for modeling the gamma-ray burst observation.
\end{itemize}

\begin{center}
\begin{tabular}{|l|c|c|p{1.5cm}|}
  \hline
Ref. & $\log_{10}(E_{k,iso})$  & $\theta_0$ & $\theta_v$  \\
  \hline
&&&\\
T & $52.47_{-0.56}^{+0.81}$  & $0.079_{-0.024}^{+0.026}$ & $0.56_{-0.16}^{+0.16}$  \\
&&&\\
  \hline
  &&&\\
Tgw & $52.80_{-0.65}^{+0.89}$ & $0.059_{-0.017}^{+0.017}$ & $0.38_{-0.11}^{+0.11}$  \\
&&&\\
  \hline
&&&\\
M, G & $52.4_{-0.7}^{+0.6}$  & $0.059_{-0.017}^{+0.017}$ & $0.26_{-0.017}^{+0.026}$ \\
&&&\\
\hline
  &&&\\
H & -\,-  & -\,- & $0.29_{-0.01}^{+0.02}$  \\
&&&\\
  \hline
  &&&\\
H & -\,-  & -\,- & $0.30_{-0.02}^{+0.02}$  \\
&&&\\
\hline
  &&&\\
W & $52.5_{-1.35}^{+1.16}$ & $0.11_{-0.02}^{+0.03}$ & $0.529_{-0.072}^{+0.129}$  \\
&&&\\
  \hline
\end{tabular}
\captionof{table}{Constraints on the jet parameters. The abbreviations used for the references are the following: T = \citep{afterglow2018} using the afterglow alone, Tgw =  \citep{afterglow2018} including the LIGO constraints on the inclination angle using the Planck value of $H_0$, M = \citep{MooleyDellerGottliebEtAl2018}, G = \citep{GhirlandaSalafiaParagiEtAl2018}, H = \citep{HotokezakaNakarGottliebEtAl2018}, W = \citep{WuMacFadyen2019}. Note that the reference H appears twice, depending on whether the jet model is gaussian or power-law. }
  \label{parcons}
\end{center}

\section{Expansion of the pick-up ring against the wind \label{B}}

In this appendix, we estimate the expansion factor of the pick-up ring in the conversion-diffusion model, after it expands against the wind under its internal pressure. To do so, we first evaluate the internal pressure at a cylindrical radius $x$ in the pick-up ring before expansion, where the Lorentz factor is $\Gamma(x)$. Due to neutron pick-up, the pick-up ring is radiation dominated, with a pressure given by

\begin{equation}
\label{Pcs} P_r(x,y) = \frac{1}{3} \frac{\delta e_{int}}{\delta V_{ring frame}} ,
\end{equation}  
in the local inertial frame, where $\delta e_{int}$ is given by Eq. \eqref{eintbis}.
Note that the transverse pressure of the ring does not depend on the frame because the ring moves radially. We consider the relativistic component for which $\Gamma(x)^2 \gg 1$, such that:

\begin{align}
\label{Pcsbis} P_r(x,y) & \sim \frac{2\Gamma_j}{9\Gamma(x)^2}y^{-2} n_{j,0} m_pc^2\\
& \sim 6\times 10^{23} \mathrm{erg} \, \mathrm{cm}^{-3} \, \frac{\Gamma_j}{\Gamma(x)^2} y^{-2} \,\, L_{j,iso,52} \Gamma_{j,2}^{-1} r_{0,7}^{-2} \, ,
\end{align} 
where we used that $\delta N_{j,r}/\delta V_{ring frame} = n_{j,0}/\Gamma(x)$. Note that we also assumed $\Gamma_j \gg \Gamma(x)$, which gives an upper bound on $\delta e_{int}$ in Eq. \eqref{eintbis} and is therefore a conservative choice for evaluating the expansion of the pick-up ring against the wind. \\
\indent As for the internal pressure in the wind which is itself radiation-dominated, it reads:

\begin{align}
\label{Pw} P_w(y) & = \frac{1}{3} a T_w^4 \sim \frac{1}{3} a T_Q^4 y^{-1} \\
& \sim 2 \times 10^{23} \mathrm{erg} \, \mathrm{cm}^{-3} \, y^{-1} \,\, \dot{M}_{32} r_{0,7}^{-1} \left(\frac{\xi}{1.5}\right) \left(\frac{\beta_{w}}{0.1}\right)^{-2} \dot{Q}_{0,19} , 
\end{align}
where $T_w$ in Eq. \eqref{TwQ} is the wind temperature and $T_Q$ is given by Eq. \eqref{TQ}. \\
\indent Finally, we can estimate the local expansion factor $g_{exp}(x)$ in the pick-up ring by requiring that after expansion, the internal pressure should be the same at every $x$ and equal to the wind pressure. As argued in Appendix \ref{sec:checks}, the internal energy evolves proportionally to the inverse area as the ring expands. Supposing that the angular size of the ring remains much smaller than $\theta_0$ (which should be eventually checked), the expansion is essentially one-dimensional, such that:

\begin{align}
\label{fexpx} g_{exp}(x,y) & = \frac{P_r(x,y)}{P_w(x)} \\
& \sim 3\times 10^2 \, \frac{1}{\Gamma(x)^2}y^{-1} \\
&\nonumber \hphantom{=} L_{j,iso,52} \dot{M}_{32}^{-1} r_{0,7}^{-1} \left(\frac{\xi}{1.5}\right)^{-1} \left(\frac{\beta_{w}}{0.1}\right)^{2} \dot{Q}_{0,19}^{-1} \, .
\end{align}
And the expansion factor $f_{exp}$ for the total pick-up ring reads:

\begin{align}
\label{fexp} f_{exp}(y) & = \frac{1}{L_{r,i}} \int^{L_{r,i}} \mathrm{d}x \, g_{exp}(x,y) \\
\nonumber & = \left<g_{exp}(x,y)\right>_{r,i} \\
\label{fexpbis} & \sim 3\times 10^2 \, \left<\frac{1}{\Gamma(x)^2}\right>_{r,i} y^{-1} \\
\nonumber & \hphantom{=} L_{j,iso,52} \dot{M}_{32}^{-1} r_{0,7}^{-1} \left(\frac{\xi}{1.5}\right)^{-1} \left(\frac{\beta_{w}}{0.1}\right)^{2} \dot{Q}_{0,19}^{-1} \, , 
\end{align}
where $L_{r,i}$ is the initial size of the pick-up ring, the integral is performed over the initial pick-up ring, as well as the average. We then use that the Lorentz factor profile in the initial pick-up ring is typically given by Eq. \eqref{Gfit} to find that:

\begin{equation}
\label{meanGm2} \left<\frac{1}{\Gamma(x)^2}\right>_{r,i} \sim \frac{\log(\Gamma_j^2)}{\Gamma_j^2},
\end{equation}
so that for $\Gamma_j \gtrsim 30$, the pick-up ring as a whole typically does not expand against the wind before breakout. Note that, from Eq. \eqref{fexpx}, sub-rings with $\Gamma(x) \lesssim 10$ are expected to expand. We check in Appendix \ref{sec:checks} that it does not prevent the ring from expanding after breakout.

\section{Consistency checks}
\label{sec:checks}

Here we check that the resulting structure for the pick-up ring is consistent with the results and hypotheses from the previous sections. In order to do so, we should make sure that on the one hand it is not in contradiction with our results for the baryon loading resulting from neutron leakage, and on the other hand that the ring is able to expand after breakout from the ejecta. 

\subsubsection*{Baryon number}

In this model of diffusion, the pick-up ring should be composed of a non-relativistic component near the border with the wind where incoming neutrons scatter elastically, followed inwards by the relativistic component where the conversion-diffusion process takes place. \\
\indent For a consistent model, we need to check that the number of baryons picked-up in the relativistic part $\delta N_{pu,r}$ is negligible compared to the total number given by Eq. \eqref{dNpuref}. This is estimated as an integral over the cylindrical radius:

\begin{equation}
\begin{split}
\delta N_{pu,r}(y) & = \int_{x_i}^{x_f} n_{b,r}(x)2\pi y r_0^2 \theta_0 \delta y \mathrm{d}x \, ,  \\ 
& \sim  \int_{x_i}^{x_f} \frac{\Gamma_j}{2\Gamma(x)^2}2\pi n_{j,0} y^{-1} r_0^2 \theta_0 \delta y \mathrm{d}x \, , \\  
& \sim \frac{3}{5} \int_{\Gamma_0}^{\Gamma_j}\frac{\mathrm{d}\Gamma}{\Gamma} \; 2 \pi r_0^2\theta_0 \sigma_{inel}^{-1} y\delta y \, , \\  
& \sim \frac{6\pi\theta_0r_0^2}{5\sigma_{inel}}\log(\Gamma_j) y\delta y \, , 
\end{split}
\end{equation}
where the integral is performed for convenience in the lab frame, but the value does not depend on the frame. We used Eq. \eqref{dNpus} in the limit $\Gamma_j^2\gg \Gamma(x)^2 \gg 1$ between the first and the second line, and Eqs. \eqref{Gld} and \eqref{ldxy} to go from the second to the third line. At $y_{r-pro}$, we finally get:

\begin{equation}
\begin{split}
\delta N_{pu,r}(y_{r-pro}) &\sim 4\times 10^{41} \; \delta y \; r_{0,7} \theta_{0,-1} \\
\nonumber &\hphantom{=} \left(1+\frac{\log(\Gamma_{j,3})}{3\log(10)}\right) \left(\frac{\beta_{w}}{0.1}\right) \left(\frac{\Delta t_{r-pro}}{0.2s}\right) \, , 
\end{split}
\end{equation} 
which is indeed much smaller than the estimated total number of baryons that drift into the jet from the wind in Eq. \eqref{dNpuref}. As expected, the majority of picked-up baryons goes into the non-relativistic elastically-scattering-neutron component, which is not relevant for the isotropic equivalent energy structure. Interestingly, as long as the total number of picked-up baryons is sufficiently larger than $\delta N_{pu,r}(y_{r-pro})$, the resulting Lorentz factor and isotropic equivalent energy structure will not depend on this number. This means in particular that the precise composition of the wind and its initial dynamics do not have an influence on the observed structure in this model (apart from 
the \textit{r}-process duration $\Delta t_{r-pro}$).

\subsubsection*{Lateral expansion before breakout}

For each part of the pick-up ring with Lorentz factor $\Gamma(x)$ to expand to an angle $1/\Gamma$ after the breakout from the ejecta, the sound speed $\sqrt{\frac{\partial p}{\partial \rho}}$ - with $\rho$ the mass density - should be sufficiently close to its relativistic value $\sqrt{\frac{1}{3}}$. The condition is for the internal energy density to be larger than the inertial mass energy density. In the single shock approximation (equations \eqref{dNpusbis} and \eqref{eintbis}), the internal energy per unit volume $u$ of a part of the pick-up ring with Lorentz factor $\Gamma(x)$ is given by:

\begin{align}
u &\sim \frac{\delta N_{j,r} m_pc^2}{\delta V_{ring frame}} \left( \frac{\Gamma_j}{\Gamma(x)}-1 \right. \\
\nonumber &\hphantom{=} \left. + \left(\frac{\Gamma_j}{\Gamma(x)}-\frac{\Gamma_j}{\Gamma(x+\lambda_{inel})}\right) \frac{\sqrt{\frac{1 - \Gamma_j^{-2}}{1-\Gamma(x)^{-2}}} - 1 }{1-\sqrt{\frac{1 - \Gamma(x+\lambda_{inel})^{-2}}{1-\Gamma(x)^{-2}}} } \right) \, , \\
&\sim \frac{\delta N_{j,r}m_pc^2}{\delta V_{ring frame}} \frac{2\Gamma_j}{3\Gamma(x)} \, , 
\end{align}
where we assumed $\Gamma_j \gg \Gamma(x)$ and $\Gamma(x+\lambda_{inel})^2\gg 1$ (we do not consider the sub-relativistic pick-up ring which is anyway irrelevant for the structure). Using Eq. \eqref{dNpusbis} with the same assumptions, we have before the pre-breakout expansion:

\begin{equation}
\label{usem}\frac{u}{n_{b,r} m_{p}c^2} \sim  1 \, , 
\end{equation}
where $n_{b,r} = \delta N_{pu,r} / \delta V_{ring frame}$. After the pre-breakout lateral expansion, the radiation energy density of the radiation dominated ring changes like $u_{f} \propto (\Sigma_{f} / \Sigma_{i}) ^{-1} u_{i}$ (see for example equation (13) in \citep{0004-637X-740-2-100}), where $\Sigma$ is the cross section of the ring, and indexes $i$ and $f$ refer to the initial and final quantities, respectively.

By conservation of the baryon number, the ratio of the initial baryon density over the final one is also inversely proportional to the ratio of the
cross sections,
so that $u/(n_{b,r}m_{p}c^2) \sim 1$ is the same before and after expansion, as long as the expanding ring is collimated by the wind. Each part of the pick-up ring can therefore expand to an angle $\sim 1/\Gamma(x)$ after the ring breaks out of the ejecta.

\subsubsection*{Timescale constraints}

\begin{figure}
\begin{center}
\includegraphics[scale = 0.5]{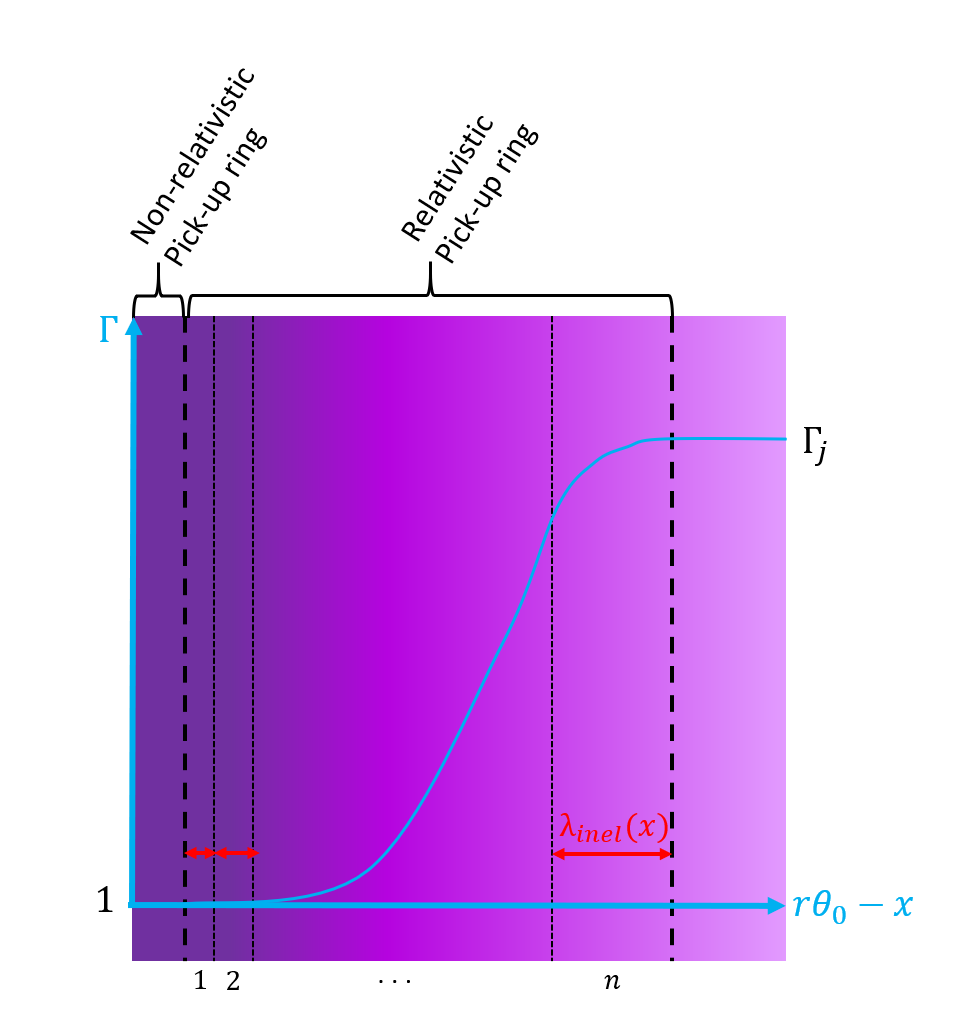}
\caption{Schematic representation of the pick-up ring in the conversion-diffusion model. The pick-up ring is divided into the non-relativistic pick-up ring near the boundary with the wind - where incoming neutrons scatter elastically - and the relativistic pick-up ring where conversion diffusion happens. We divide the latter into sub-rings labeled $1,2,\dots ,n$ at radii $x_1, x_2,\dots , x_n$ starting from the outer edge, and such that $x_{i+1} = x_i - \lambda_{inel}(x_i)$.}
\label{fig:pu_ring}
\end{center}
\end{figure}

We finally check that the conversion-diffusion process is not altered by timescale considerations, corresponding to the diffusion time being smaller than the dynamical time. 
Usually the diffusion time to cross a distance $n \lambda_{inel}(x)$ is $\sim n^2 (\lambda_{inel}(x)/c\beta)$ where 
$\lambda_{inel}(x)$ is the mean free path.
However, it is not $\propto n^2$ (diffusive) but close to $\propto n$ (ballistic-like) in our problem. The argument goes as follows.

\begin{figure}
\begin{center}
\includegraphics[scale = 0.6]{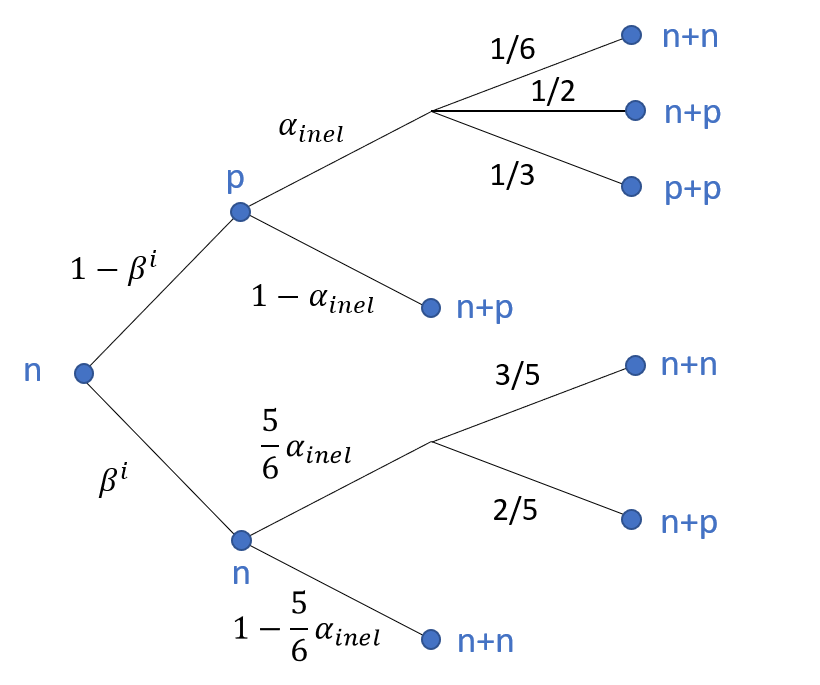}
\caption{Probability tree in ring  $i$, 
summarizing the possible interactions between a neutron coming 
from ring $i-1$ (located to the left of the figure) and a nucleon 
in ring $i$. In blue are indicated the reacting and product particles, and in black the probability of each branch. Whereas the first branches correspond to the choice of the nucleon interacting with the incoming neutron, the following 7 branches indicate the probability for each product. }
\label{fig:scatree}
\end{center}
\end{figure}
\begin{figure}
\begin{center}
\includegraphics[scale = 0.5]{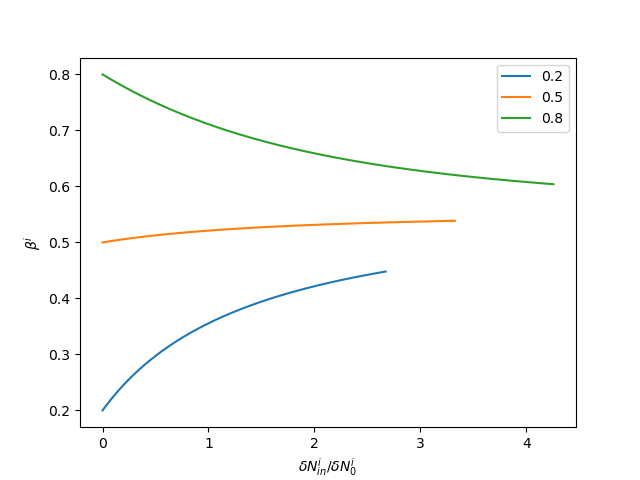}
\caption{Evolution of the neutron fraction in ring $i$, $\beta^i$, as a function of $\delta N_{in}^i$ from the solution of Eqs. \eqref{ODENpu}-\eqref{ODENin} for $\alpha_{inel} = 0.5$, $\delta N_0^i = 10^{38} \delta y$ and various initial values of $\beta^i$ indicated in the legend for each curve. We end the integration when $\delta N_{pu}^i = \delta N_0^i$, which happens before the end of pick-up for the region of the pick-up ring relevant to the jet's structure (where $\Gamma \lesssim 10$).}
\label{beta}
\end{center}
\end{figure}

 We consider the situation illustrated in Fig. \ref{fig:pu_ring}: the relativistic pick-up ring (where the conversion-diffusion happens) is divided into sub-rings labeled $1,2,\dots ,n$ at radii $x_1, x_2,\dots , x_n$ starting from the outer edge, and such that $x_{i+1} = x_i - \lambda_{inel}(x_i)$. We further define $\delta N_0^i$ the number of baryons initially in the $i$-th ring, $\delta N_{pu}^i$ the number of neutrons picked-up in the $i$-th ring, $\beta^i$ the neutron fraction in the $i$-th ring and $\delta N_{in}^i$ the number of incoming neutrons at the $i$-th ring. The possible interactions of a neutron coming from ring $i-1$ with nucleons of ring $i$ are 
 \begin{align}
\nonumber n+p &\to  n+p \, , \, \mathrm{(elastic)\, with\, cross-section\,} \sigma_{el,p}\\
\nonumber n+p &\to  n+p+\pi^0 \, , \, \mathrm{with\, cross-section\,} \sigma_{0,p}\\
\nonumber n+p &\to  n+n+\pi^+ \, , \, \mathrm{with\, cross-section\,} \sigma_+\\
\nonumber n+p &\to  p+p+\pi^- \, , \, \mathrm{with\, cross-section\,} \sigma_{-,p}\\
\nonumber n+n &\to  n+n \, ,\, \mathrm{(elastic)\, with\, cross-section\,} \sigma_{el,n}\\
\nonumber n+n &\to  n+n+\pi^0 \, ,\, \mathrm{with\, cross-section\,} \sigma_{0,n}\\
\nonumber n+n &\to  n+p+\pi^- \, , \, \mathrm{with\, cross-section\,}  \sigma_{-,n} \, ,
\end{align}
and we further define the total inelastic cross-sections
\begin{align}
\sigma_{inel,p} &\equiv \sigma_{0,p} + \sigma_+ + \sigma_{-,p} \, , \\
\sigma_{inel,n} &\equiv \sigma_{0,n} + \sigma_{-,n} \, .
\end{align}
As the precise experimental data for nucleon-nucleon scattering in the inelastic regime is not available, we assume for concreteness:
\begin{align}
\sigma_{el,p} &= \sigma_{el,n} \equiv \sigma_{el} \, , \\
\sigma_{0,p} &= \sigma_{0,n} \equiv \sigma_0 \, , \\
\sigma_{-,p} &= \sigma_{-,n} \equiv \sigma_- \, , \\
\sigma_0 &= \sigma_- + \sigma_+ \, .  
\end{align}
These assumptions are expected to be reasonable due to approximate isospin symmetry of the strong interactions. Again for concreteness, we assume that
\begin{equation}
\sigma_+ = \frac{1}{2} \sigma_- \, ,
\end{equation}
which implies that the total inelastic cross-sections for $(n,p)$ and $(n,n)$ scatterings are such that
\begin{equation}
\sigma_{inel,n} = \frac{5}{6} \sigma_{inel,p} \, .
\end{equation}
Then, if we denote $\alpha_{inel}$\footnote{$\alpha_{inel}$ is an increasing function of $\Gamma_{rel}$. In particular, for $\Gamma_{rel}=\gamma_{el}$, $\alpha_{inel}=1/2$, where $\gamma_{el}$ is defined in Eq. \eqref{gel}.} the probability for an $(n,p)$ scattering to be inelastic, the probability tree is given by Fig. \ref{fig:scatree}. This yields in turn the following system of differential equations for the baryon numbers in the pick-up ring:
\begin{eqnarray}
\label{ODENpu}  \frac{\mathrm{d}\delta N_{pu}^i}{\mathrm{d}\delta N_{in}^i} &=& (1-\beta^i)\left(\frac{1-\alpha_{inel}}{2} + \frac{7}{12}\alpha_{inel}\right) \\
\nonumber &\hphantom{=}& + \frac{1}{6}\beta^i\alpha_{inel} \, , \\   
\label{ODEbeta}  \frac{\mathrm{d}\beta^i}{\mathrm{d}\delta N_{in}^i} &=& \frac{1}{6(\delta N_{pu}^i + \delta N_0^i)} \\
\nonumber &\hphantom{=}& \left((1-\beta^i)\left(3 - \frac{\alpha_{inel}}{2} - \beta^i \left( 3 + \frac{\alpha_{inel}}{2}\right) \right) \right. \\
\nonumber &\hphantom{=}& \left. - \beta^i (1+\beta^i)\alpha_{inel}\right) \, , \\
\label{ODENin}  \frac{\mathrm{d}\delta N_{in}^{i+1}}{\mathrm{d}\delta N_{in}^i} &=& (1-\beta^i)\left(\frac{1-\alpha_{inel}}{2} + \frac{5}{12}\alpha_{inel}\right) \\
\nonumber &\hphantom{=}& + \beta^i\left(\frac{2}{3}\alpha_{inel} + 1-\frac{5}{6}\alpha_{inel}\right) \, .
\end{eqnarray}
where we assume for simplicity that $\alpha_{inel}$ can be treated as a constant, equal to its final value. This assumption is justified by the fact that most neutrons are captured when $\Gamma_{rel}$ (and therefore $\alpha_{inel}$) is close to its final value.  In each reaction, half of the emitted neutrons are scattered away from the jet's axis. These back-scattered neutrons are considered to be picked-up, as they are targets for the incoming neutrons. They could in principle leave the $i$-th ring and reach inner regions of the jet on diffusive timescales, so that $\delta N_{in}^{i+1}$ and $\delta N_{pu}^i$ may receive corrections from the diffusive exchange of neutrons within the pick-up ring. As can be seen on Fig. \ref{beta} though, the neutron fraction in the pick-up ring (after pick-up is over) is expected to be roughly equal to the proton fraction, so that the diffusion of back-scattered neutrons is not expected to modify much the pick-up ring structure. Also, the further diffusion of neutrons actually makes conversion-diffusion more efficient, so that ignoring such diffusion is a conservative approximation. This is the reason why we consider the conversion-diffusion to be essentially ballistic-like. 

As we showed in appendix \ref{B} that the pick-up ring does not expand under its internal pressure against the wind before breakout if $\Gamma_j \gtrsim 30$, the diffusion time at cylindrical radius $x$ reads:

\begin{equation}
\label{Tdconvdiff} \tau_d^{c-d} = \int \frac{\mathrm{d}x}{v_{\perp}(x)} \sim \frac{\lambda_{pu}^{c-d}\left<\Gamma\right>}{c} ,
\end{equation}
where $\lambda_{pu}^{c-d}$ is the typical size of the pick-up ring (the superscript ``c-d" referring to conversion-diffusion to distinguish from the elastic scattering case) and the average of the Lorentz factor is over the pick-up ring. Imposing that the diffusion time is smaller than the dynamical time in Eq. \eqref{tdyn} gives a constraint on the size of the pick-up ring:

\begin{equation}
\label{bound} \frac{\lambda_{pu}^{c-d}}{r} < \frac{1}{\left<\Gamma\right>} ,
\end{equation}
and using that the Lorentz factor profile in the pick-up ring is typically given by Eq. \eqref{Gfit}:

\begin{equation}
\frac{\lambda_{pu}^{c-d}(y)}{r} \sim \frac{3\Gamma_j y}{5n_{j,0}r_0\sigma_{inel}} , 
\end{equation} 
\begin{equation}
\left<\Gamma\right> \sim \frac{2}{3} \left(\lambda_{pu}^{c-d}\right)^{1/2} \left(\frac{5}{3}\Gamma_j n_{j,0}y^{-2}\sigma_{inel}\right)^{1/2} \sim \frac{2}{3} \Gamma_j,    
\end{equation}
so that
\begin{align}
\label{dyncons}\frac{\lambda_{pu}^{c-d}(y)\left<\Gamma\right>(y)}{r} &\sim \frac{2\Gamma_j^2 y}{5n_{j,0}r_0\sigma_{inel}} \\
&\sim 5\times 10^{-3} \, y\, \Gamma_{j,3}^3 L_{j,iso,52}^{-1} r_{0,7} .
\end{align}
The latter quantity is upper-bounded by the value at the $r-$process radius $y_{r-pro}$:
\begin{equation}
\label{dynconsrpro}\frac{\lambda_{pu}^{c-d}\left<\Gamma\right>}{r}(y_{r-pro}) \sim 2 \, \Gamma_{j,3}^3 L_{j,iso,52}^{-1} \left(\frac{\beta_w}{0.1}\right)\left(\frac{\Delta t_{r-pro}}{1\,\mathrm{s}}\right) ,     
\end{equation}
so that the condition \eqref{bound} is obeyed and the diffusion time may be smaller than the dynamical time for typical parameters.

\end{document}